\documentclass[fleqn,usenatbib]{mnras}

\usepackage[T1]{fontenc}
\usepackage{ae,aecompl}
\usepackage[dvipsnames]{xcolor} 

\usepackage{booktabs}
\usepackage{array}
\usepackage{siunitx}
\usepackage{graphicx}	

\usepackage{newtxtext,newtxmath}

\usepackage{amsmath}	
\usepackage{amssymb}	
\usepackage{cleveref}
\crefformat{section}{\S#2#1#3} 
\crefformat{subsection}{\S#2#1#3}
\crefformat{subsubsection}{\S#2#1#3}

\newcolumntype{L}{>{$}l<{$}}
\newcolumntype{C}{>{$}c<{$}}
\newcolumntype{R}{>{$}r<{$}}

\def\simeq{
\mathrel{\raise.3ex\hbox{$\sim$}\mkern-14mu\lower0.4ex\hbox{$-$}}
}

\def\msun{{\rm M_{\odot}}}
 
\def\le{{L_{\rm Edd}}}

\def\del#1{{}}
\def\ltsima{$\; \buildrel < \over \sim \;$}
\def\simlt{\lower.5ex\hbox{\ltsima}}
\def\gtsima{$\; \buildrel > \over \sim \;$}
\def\simgt{\lower.5ex\hbox{\gtsima}}

\def\deg{^{\circ}}

\title[Improving Black Hole Accretion Treatment]{Improving Black Hole Accretion Treatment in Hydrodynamical Simulations}

\author[M. Tart\.{e}nas \& K. Zubovas]{
Matas Tart\.{e}nas$^{1}$\thanks{E-mail: matas.tartenas@ftmc.lt},
Kastytis Zubovas$^{1,2}$
\\
  $^{1}$Center for Physical Sciences and Technology, Saul\.{e}tekio av. 3, Vilnius LT-10257, Lithuania \\
  $^{2}$Astronomical Observatory, Vilnius University, Saul\.{e}tekio av. 3, Vilnius LT-10257, Lithuania\\
}

\date{Accepted XXX. Received YYY; in original form ZZZ}

\pubyear{2021}

\begin{document}
\label{firstpage}
\pagerange{\pageref{firstpage}--\pageref{lastpage}}
\maketitle

\begin{abstract}
The large galactic scales are connected to the many orders of magnitude smaller supermassive black hole (SMBH) scales by an episodic cycle of feeding and feedback. Active galactic nuclei (AGN) are powered by accretion onto SMBH and the majority of AGN energy, in near-Eddington regime, is produced in thin sub-pc accretion discs.  Currently, it is very difficult to model processes that occur on vastly different scales, ranging from the circumnuclear gas reservoirs at tens to hundreds of parsecs, down to the accretion disc scales at <0.01 pc. While sub-grid prescriptions used in large-scale or cosmological simulations are able to reproduce large-scale feedback, we propose using a more realistic model in parsec-scale simulations, where it is important to get accurate timescales to understand how feedback affects gas dynamics and star formation in the vicinity of the AGN. To test our approach we use a sub-resolution thin accretion disc model, coupled to the SMBH, in a set of hydrodynamical simulations of a retrograde collision between a gas ring and a molecular cloud in an environment similar to the Galactic centre using the SPH code Gadget-3. The disc-mediated feeding of the SMBH is relatively smooth and delayed compared to an instantaneous feeding prescription. While the reduction of accretion due to feedback is present in both accretion disc and instantaneous feeding simulations, a clear central cavity appears only in accretion disc runs - hinting that a less volatile accretion phase could have a greater impact on the surrounding gas.
\end{abstract}




\section{Introduction} \label{intro}

Various correlations between properties of the supermassive black holes (SMBH) found at the centres of galaxies with the properties of their respective host galaxies imply a close link between galaxy and SMBH evolution \citep[eg.][]{1998Rees, King2003, Ciotti2007a, Ciotti2007b, 2009Cattaneo, Novak2011, 2013Kormendy, Ciotti2017}. Understanding how this co-evolution occurs is an important challenge that requires both observation and increasingly detailed modeling.

Much of the difficulty in understanding these systems comes from the fact that SMBH and relevant scales (sub-parsec to a few parsecs) are extremely small when compared with the galactic (kilo-parsec) scales, yet we see that the large and the small galactic scales are connected by an episodic cycle of feeding and feedback \citep{Gaspari2020}. Kilo-parsec scale outflows are detected in many galaxies \citep{McKinley2021,Laha_rew2021} and observations at pc-scale resolution reveal an intricate picture of feeding and feedback in currently active galaxies \citep[eg.][]{Burillo2005,NGC10682019, NGC6132019, NGC10972019, General_circinus_NGC1068, NGC1275_2019, Burillo2021}. Some of these features are also detected in our own Galaxy, which is currently inactive \citep{gcrev}. Large, kpc-scale, outflows \citep{fermior, eRosita2020} are linked to the Galactic Centre by a $\sim100$~pc bridge of X-ray chimneys and radio bubbles \citep{XRAYCHIMNEY, 430RADIOBUBBLES}.  

Cosmological and galactic-scale simulations are capable of recreating the observed scaling relations when feedback is included \citep[eg.][]{Di_Matteo_2005,Filloux_2010,iliustris2014, Steinborn_2015, Eagle22015, Alcazar2016, Romulus2017,TNG12019, TNG22019}. In addition, the inclusion of feedback allows the simulations to recreate and explore the observed features such as Fermi bubble-like outflows \citep{TNG50_bubbles2021} or the impact of active galactic nuclei (AGN) on the interstellar medium (ISM) \citep{SIMBA_2019, ANG_wind_impact2020}. While these models are relatively successful at recreating the large scale features and observed relations, they are not capable, nor try to, accurately model the parsec-scale gas dynamics.
In order to study SMBH feeding, AGN feedback and their impact on the local environment in more detail we still need to refer to smaller-scale simulations \citep[eg.][]{StarFormationGC1, Alig2, LUCASMCINFALL,Hopkins2016, CMZMOD, Tartenas2020, Tress2020} where gas dynamics in central parts of a galaxy can be resolved or utilize  sub-resolution prescriptions to track unresolved physical processes \citep{Negri_2017}.

The usual way to circumvent the limitation of finite resolution in large numerical simulations is to use a sub-grid prescription for an unresolved physical process. A go-to sub-grid prescription for the process of accretion was proposed by \cite{Springel_BONDi_2005} and uses the Bondi-Hoyle-Lyttleton accretion flow solution \citep[Bondi method;][]{Bondi_Lyttleton_1939, Bondi_hoyle_1944, Bondi_1952}. This method is very convenient, especially in large scale simulations, as the SMBH feeding depends entirely on gas properties ``far away'' from the black hole. In this case, gas particles are not captured and instantly removed, but are only used to determine said ambient properties on which accretion depends. Once the SMBH accretion rate is determined, its mass is increased and particles are removed stochastically from the surrounding medium ensuring approximate conservation of mass. However, the classical Bondi approach does not provide a unique way to evaluate the ambient gas properties. It also requires the unrealistic assumption of absence of angular momentum which may lead to an over- or underestimation of SMBH accretion in certain situations \citep{Hobbs_2012,Negri_2017}. These issues are somewhat mitigated by introducing a numerical correction factor $\alpha$ \citep{Springel_BONDi_2005} and/or using other modifications to the Bondi method \citep[e.g.,][]{Booth2009,AngMomInBondi, Eagle1}. An alternative to the Bondi method that also depends on the ambient properties of gas uses gravitational torques within a certain radius of the SMBH to determine its rate of accretion \citep{Alcazar2016}.

Another approach is to calculate the amount of SMBH accretion from the amount of matter added to the accreting sink particle. This amount is determined not by the ambient properties of gas surrounding the sink particle (or at least not directly), but only by the particles that get swallowed after coming "close enough" and/or fulfilling some other accretion criteria (eg. low angular momentum, low internal energy, etc.). This method can be extended by coupling the SMBH to a gas reservoir that stores the accreted material and `drip-feeds' it to the SMBH \citep{Power2011, GizmoMethod_2015}. These ``two-stage'' prescriptions allow the simulators to delay and keep the black hole accretion rate close to some chosen accretion disc model. The main disadvantage of this sort of a two-stage prescription is that the properties of the accretion flow are largely dependent on a set of freely chosen parameters - viscosity timescale, accretion efficiency, etc. In particular, the viscous timescale varies dramatically even in an $\alpha-$prescription accretion disc depending on the assumed accretion radius ($t_{\rm visc}\propto R^{3/2}$). This means that this single parameter may artificially increase or reduce the amount and significantly impact the timing of feedback injections in a given simulation, since $t_{\rm visc}$ might vary between a few yr and a few Myr inside a given accretion disc \citep{Tartenas2020}, although in practice some `reasonable number' in between the two extremes is often chosen. This may not be as important in large scale simulations, but it may be critical if we want to, for example, isolate the impact that black hole wind has on star formation in the central $\sim100$~pc. Results from small scale simulations also help to improve existing sub-grid prescriptions used in large scale simulations.

In this paper, we aim to improve the tracking of SMBH accretion in small-scale simulations. To do so, we extend the usual two-stage SMBH accretion model for hydrodynamical simulations by including a simple $\alpha$-accretion disc \citep{ShakuraSunyaev} prescription coupled to the SMBH sink particle. This allows us to reduce the number of free parameters with a relatively low impact on computational cost. The disc is evolved viscously with a separate timestep criterion from the simulation as a whole, but is synchronized at SMBH timesteps. Feedback is directly determined from the parameters of the accretion disc, with radiation efficiency naturally aproaching $\eta\approx0.0625$, which is correct for our chosen Paczy\'nsky-Wiita potential \citep{Paczy1980}. We test the prescription in a smoothed-particle hydrodynamics (SPH) \texttt{Gadget-3} simulation \citep{GADGET2005} of a cloud impacting a gas ring around the SMBH. We show that feedback has a significant effect in regulating the growth of the SMBH. Models with the accretion disc method produce a clear cavity in the centre, abruptly stopping any further accretion; this is not reproduced by models with instantaneous accretion. We argue that our result is more realistic than the alternative.

The layout of this paper is as follows. First, we describe the the general setup of our simulations in \cref{sec:PaN}. Section \cref{sec:AccPartSetup} shows the key assumptions and equations determining the evolution the accretion disc particle. Results of the main set of hydrodynamical simulations are shown in \cref{results}, with possible implications, limitations and further research directions discussed in \cref{Discussion}. Conclusions follow in \cref{sec:Conclusions}.

\section{Numerical setup}\label{sec:PaN}

As our accretion disc method is intended to improve simulations of the vicinity of an AGN, we test our approach with a set of simulations of retrograde collisions between a gas ring and a molecular cloud in an environment similar to that of the Milky Way centre. In a previous paper \citep{Tartenas2020} we showed that similar configurations, without feedback, result in an AGN phase lasting $\gtrsim100$~kyr.

We use the N-body/SPH code \texttt{Gadget-3} \citep{GADGET2005} with the SPHS formulation \citep{SPHS2012} and the appropriate Wendland kernel function C$^2$ \citep{kernel} with neighbour number $N_{\rm neigh} = 100$\footnote{In the making of this paper we extensively used \texttt{matplolib} \citep{matplotlib}, \texttt{pygadgetreader} \citep{pygadgetreader}, \texttt{numba} \citep{numba} and \texttt{numpy} \citep{numpy}}. The gas ring and cloud are each composed of $N_{\rm part} \approx 5\times10^5$ particles of mass $m_{\rm SPH} \approx 0.4 \, \msun$. The resolved mass is $M_{\rm res} = N_{\rm neigh} m_{\rm SPH} \approx 40 \textrm{M}_{\odot}$. The initial velocities are determined using a potential given by:
\begin{equation}
   \phi = -\frac{GM_{\rm BH}}{r} + 2\sigma^2 \log \frac{r}{r_0},
\label{pot}
\end{equation}
where the first term is the gravitational potential of a point mass (the SMBH) and the second is an isothermal potential with velocity dispersion $\sigma = 100$~km$\,$~s$^{-1}$; r$_{0}$ is an arbitrary large constant. Our chosen potential corresponds to an enclosed mass $M_{\rm enc} = M_{\rm bh}$ at $R_{\rm enc} = 0.8$~pc. A more detailed description of the initial conditions and the physics included in our simulations is given below.

\subsection{Initial Conditions \label{sec:IC} } 

\begin{figure}
	\includegraphics[width=\columnwidth]{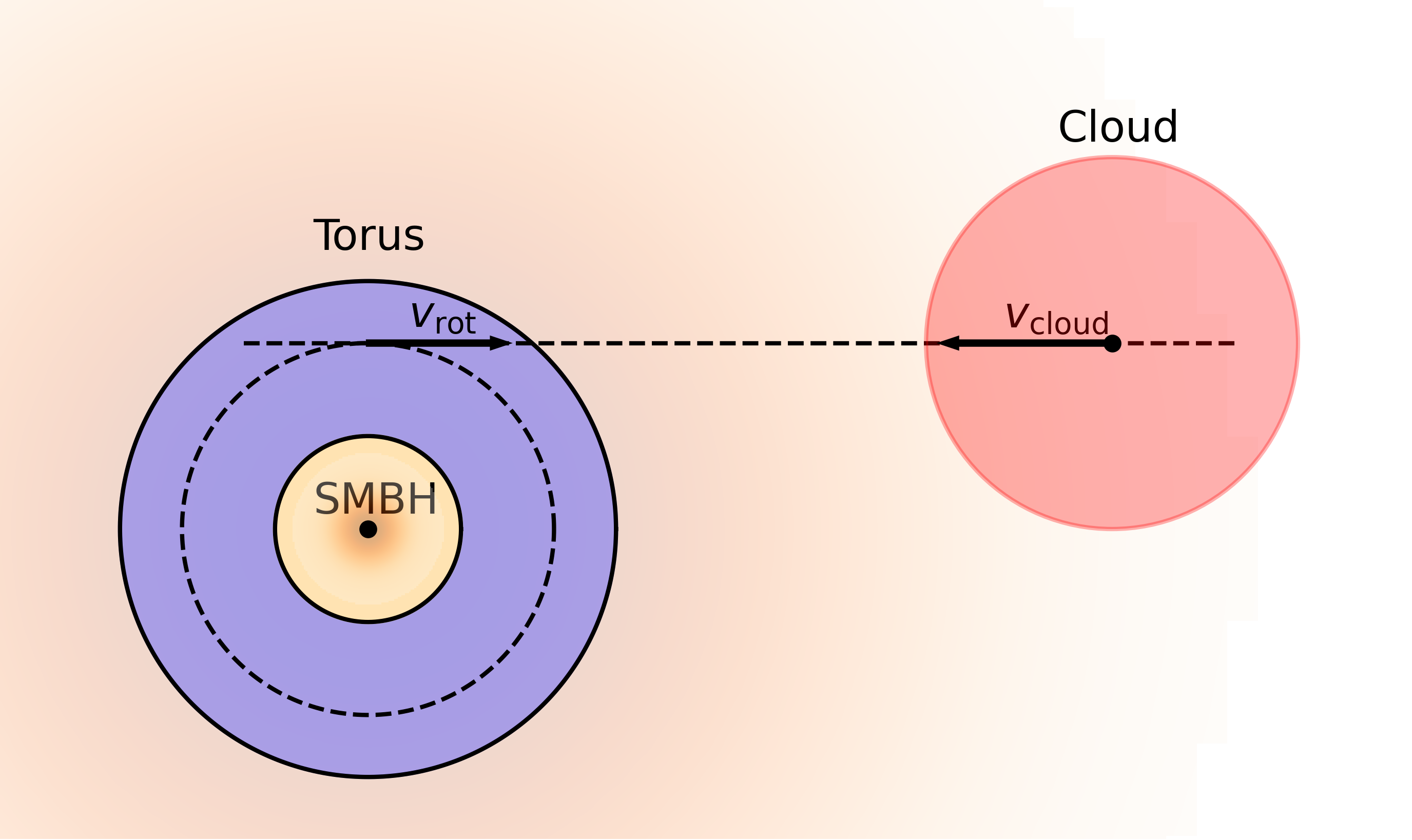}
	\vspace{-0.5cm}
    \caption{ Elements of the model as seen along the $z$-axis: CNR-like torus (blue); the infalling molecular cloud (red); the background gas (orange); the SMBH (black dot). The cloud is placed on a collision course with the torus. All elements share the $xy$-midplane.}
    \label{fig:TorCloud}
\end{figure}

\textbf{The supermassive black hole}: Our model contains a SMBH with an initial mass of $M_{\rm BH} = 4\times10^6\,\msun$. The chosen mass is similar to the SMBH mass at the centre of the Milky Way, determined from the orbits of S stars - $4.02\,\pm0.16\pm\,0.04 \times10^6\,\msun$ \citep{BHMASS2016}. The SMBH is fixed at the centre of the system. The SMBH is coupled with a sub-resolution accretion disc. Particles are accreted by this combined SMBH-accretion disc particle if they fall inside $r_{\rm sink}=0.01$~pc, which is approximately the minimal volume spatial resolution in the model, and have orbits with circularization radius $r_{\rm circ} < r_{\rm sink}$. This excludes the accretion of particles with high angular momentum. Further evolution of this gas is followed using the accretion disc particle method (see sec. \ref{sec:AccPartSetup}).

\textbf{The circumnuclear ring}: The toroidal gas ring  is similar in size and mass to the Circumnuclear Ring (CNR) found at the centre of the Milky Way \citep{minmass}. It has an inner radius $R_{\rm in} = 1.5$~pc, an outer radius $R_{\rm out} = 4$~pc. The mass estimates of the current CNR vary by almost two orders of magnitude; for simplicity we set the initial mass of the ring to $M_{\rm ring} = 10^5\, \msun$ \citep{gas} and set its density to a constant value. Ring particles move in circular orbits with speeds $v_{R1.5} \sim 181$ km s$^{-1}$ at the inner edge and $v_{R4} \sim 160$ km s$^{-1}$ at the outer edge.

\textbf{The molecular cloud}: An infalling molecular cloud is placed $12$~pc away from the target (Fig. \ref{fig:TorCloud}). The cloud contains the same amount of gas as the CNR $M_{\rm cloud} = 10^5\, \msun\,$  and is also of constant density. We set the radius $r_{\rm cloud} = 3$~pc based on observational estimates of typical sizes of clouds of this mass in the Galaxy \citep{cloudSurvey}. The cloud is set on a parabolic orbit that crosses the ring at a point $3$~pc away from the origin along the Y axis. The initial velocity of the centre of the cloud is $v_{\rm cl} = 220$km s$^{-1}$; this corresponds to a parabolic velocity at the position of the cloud's centre. While the molecular cloud in our simulations does not correspond to any particular feature currently observed in the Galactic Centre, an appearance of such a cloud over several million years seems likely given that both observations and simulations show an inflow of matter into the central few parsecs \citep{Sormani2019, Tress2020}. In fact, the CNR itself could have been formed by a capture of a similar cloud \citep{CNRVirialmass, CNRform2015, CNRform2018}.

\textbf{Background gas}: The primary intent in including the diffuse background is to facilitate the tracking of the feedback energy input, in particular in directions perpendicular to the rotation plane of the CNR. As a result, we set up only a strongly idealized spherical gas distribution. It contains $M_{\rm bg} = 1.2\times10^3 \msun$ extending out to $r_{\rm bg} = 25$~pc and following an isothermal density profile. The background gas consists of $N_{\rm bg} \approx 3\times10^5$ particles; the mass of each particle is 100 times lower than that of particles in the cloud and the ring. The initial velocity of the background gas is set to zero. 

\textbf{Turbulence}: In addition to orbital velocities, MC and CNR particles are given velocities from a turbulent velocity field. The velocity field is generated based on the example of \cite{Turbulencija}, with velocity amplitude $\sigma_{\rm turb} = 37.5$ km s$^{-1}$. Turbulence is only present initially and dissipates over time, since currently there is no further turbulence driving. Nevertheless, it results in the formation of a turbulent density field, which is more realistic than smooth gaseous structures.

\subsection{Cooling} \label{sec:Cooling}
The thermal properties of the gas particles are determined using two methods depending on the temperature of the gas. For gas with temperatures between $20$~K and $10^4$~K an empirical function by \cite{cool_to_20K} is used:
\begin{equation}
    \log(\Lambda/n^2_{\rm H})= -24.81 + 2.92x - 0.6982x^2+\log(Z/Z_{\odot}),
\end{equation}
here $x \equiv \log(\log(\log(T)))$;  $\Lambda/n^2_{\rm H}$ is in units of erg$\,$s$^{-1}$cm$^{3}$; $n_{\rm H}$ is the number density of H (in cm$^{-3}$). $\log(Z/Z_{\odot}$ is the metallicity in solar units; we assume $Z = Z_\odot$, therefore the metallicity term disappears. This prescription is based on the assumption that radiative cooling occurs via fine structure and metastable lines of C, N, O, Fe, S, and Si and that the above elements are in ionization equilibrium maintained by locally produced cosmic rays. 

For temperatures above $T \geq 10^4$~K we follow an approach from \cite{QuasarHEATING} where the change in energy $\dot{E}$ is approximated by:
\begin{equation}
    \dot{E} = n_{\rm H}^2 (S_1 + S_2 + S_3). 
    \label{eq:QH}
\end{equation}
Here, $S_1, S_2$ and $S_3$ correspond to the effect of different physical processes:
\begin{equation}
    S_1 = -3.8\times10^{-27} T^{1/2},    
\end{equation}
is the bremsstrahlung cooling term, 
\begin{equation}
    S_2 = 4.1\times10^{-35}(1.9\times10^7 - T)\xi    
\end{equation}
is the Compton heating or cooling and 
\begin{equation}
    S_3 =10^{-23} \frac{a + b ( \xi / \xi_0)^c}{1 + ( \xi / \xi_0)^c}
\end{equation}
is a combination of photoionization heating and line and recombination continuum cooling. The parameters here are:
\begin{align}
    a &= -\frac{48}{\exp({25\log{T} -4.35)^2 }}
    -\frac{80}{\exp({5.5\log{T} - 5.2)^2 }}-\\
    &-\frac{17}{\exp({3.6\log{T} -6.5)^2 }},\\
    b &= \frac{1.7\times10^4}{T^{0.7}},\\
    c &= 1.1 - \frac{1.1}{\exp(T/1.8\times10^5)} + \frac{4\times10^{15}}{T^4},\\
    \xi_0 &= \frac{1}{1.5 T^{-1/2}
    1.5\times10^{12} T^{-5/2}} + \\
    &+ \frac{4\times10^{10}}{T^2} \left[ 1 + \frac{80}{\exp( (T-10^4)/1.5\times10^3 )} \right]. \\
\end{align}

To determine the ionization parameter $\xi$ we use:
\begin{equation}
    \xi = \frac{L_{\rm{disc}}}{n r^2},
\end{equation}
where $L_{\rm{disc}}$ is the luminosity of the accretion disc. 

For temperatures $T>3\times10^7$~K, the $S_3$ term is dropped from equation (\ref{eq:QH}) \citep{QuasarHEATING}.

In addition to the above (mainly cooling) processes, a background photoelectric heating from grains is taken into account. Using functions from \cite{UVBGHEATING}, the photoelectric heating rate is given by:
\begin{equation}
    \Gamma_{\rm pe} = 10^{-24} \epsilon \chi \rm{n}_{\rm{H}}\,\rm{erg}\,\rm{s}^{-1}\,\rm{cm}^{-3},
\end{equation}
where $\epsilon$ is the heating efficiency and $\chi$ is the far-UV flux normalized to the Habing field appropriate for the Solar neighborhood; $\chi=100$ is appropriate for the Galactic Centre. An approximate expression for the heating efficiency is:
\begin{equation}
    \epsilon = \frac{4.87\times10^{-2}}{1 + 4\times10^{-3}(\chi T^{1/2}/n_e)^{0.73}} + \frac{3.65\times10^{-2}(T/10^4\rm{K})}{1+2\times10^{-4}(\chi T^{1/2}/n_e)},
\end{equation}
where $n_{\rm e}$ is the electron number density defined in terms of a ionization fraction $f_{\rm ion} = n_{\rm e} / n_{\rm gas} = 10^{-3}$. The precise value of $n_{\rm e}$ has a negligible effect on our results.

The balance of heating and cooling depends on the distance from the SMBH, the AGN luminosity and the gas density. At the spatial scales relevant for our simulations, with $L_{\rm AGN} = 0.5\le{}$, the equilibrium temperature of background gas is about $10^6$~K. The much denser torus and cloud have equilibrium temperatures of order $\sim10^5$~K. This produces a dichotomy which is useful for our purposes: the background gas is easily heated to temperatures higher than the virial temperature ($T_{\rm vir} \sim 5\times10^5$~K in our chosen potential), so the background does not accrete rapidly on to the SMBH even in the absence of feedback. The cloud and ring, on the other hand, are not heated to virial temperatures and so behave more-or-less balistically. 

\subsection{Star formation} \label{sec:star_formation}

Our current model setup lacks resolution to accurately simulate star formation in the vicinity of the SMBH (star formation near the SMBH is discussed further in section \cref{sec:lost_gas}), but the removal of overly dense particles from hydrodynamical calculations helps speed up the simulation without sacrificing realism. Therefore, we introduce the following star formation prescription. Gas particles are probabilistically transformed to collisionless star particles if their density is higher than the tidal density and their Jeans mass is lower than the resolved mass. Star formation is related to the dynamical time of the particle:
\begin{equation}
    t_{\rm{dyn}, i} = \sqrt{ \frac{3\pi}{32\rm{G}\rho_i}}
\end{equation}
which allows us to define the probability of the transformation, $P_{\rm{sf},i}$, as:
\begin{equation}
    P_{\rm{sf},i} = 1 - {\rm exp}{ \left ( -\frac{\eta_{\rm sf} \Delta t_i} {t_{\rm dyn,i}}\right) },
\end{equation}
here $\eta_{\rm sf} = 0.1$ is the star formation efficiency; $\Delta t_i$ is the timestep of the particle. 

\subsection{Feedback} \label{sec:Feedback}

The physical basis of our feedback implementation is the AGN wind feedback model \citep[for a review, see][]{King2015ARAA}. Numerically, we implement it with an extended version of the Monte Carlo radiation transfer method from \cite{MCRadTrans}. Here, the AGN wind is represented by a number of isotropically distributed discrete energy-momentum packets generated at the source location. The number of these is determined by the luminosity of the accretion disc particle. The momentum of a single packet is defined in relation to the SPH particle mass, $p_{\rm{\gamma}}=p_{\rm{sph}}=m_{\rm{sph}}\times \sigma \approx 8\times10^{39} \rm{g}\,\rm{cm}\,\rm{s}^{-1}$. Since the momentum is defined based on the higher mass of cloud/ring rather than background particles, our approach may introduce artificial shocks in the background gas. However, since the primary purpose of the background gas is as a tracer of energy imparted by the AGN, this has little impact on our results. 

An emitted packet moves steadily outwards with the velocity of $v_{\rm{\gamma}}=0.1c$. Once it comes close enough to some SPH particles it transfers its momentum and and energy to them over several steps. The SPH density field is directly used to determine the amount of momentum passed to each particle:
\begin{equation}
    \Delta \textbf{p}_{\rm{\gamma},i}= \frac{\rho_{i}(\textbf{r})}{\rho(\textbf{r})}\Delta \textbf{p}_{\rm{\gamma}}
\end{equation}
here $\Delta \textbf{p}_{\rm{\gamma}}$ and $\Delta \textbf{p}_{\rm{\gamma}, i}$ are the total amount of momentum transferred to the gas in a given step and the momentum transferred to the $i$-th particle, while $\rho(\textbf{r})$ and $\rho_i(\textbf{r})$ are the density field value at that point and the contribution of the $i$-th particle to that field. The radiation pressure force $\textbf{f}_{\rm{rad}, i}$ on each particle then is:
\begin{equation}
    \textbf{f}_{\rm{rad}, i} = \frac{\sum_\gamma \Delta \textbf{p}_{\rm{\gamma},i}}{\Delta t_i}.
\end{equation}
The same principle is applied when calculating the energy transfer, except that we replace $\Delta \textbf{p}_{\rm{\gamma}}$ with $\Delta E_{\rm{\gamma}}$, where $E_{\rm{\gamma}} = \eta p_{\rm{\gamma}}c /2$, where the radiative efficiency $\eta= L_{\rm disc} / \dot{M}_{BH} c^2 $ is determined directly from the accretion disc model. The total energy injected to the gas approaches \citep{Ring_Eta}
\begin{equation}
   E_{\rm wind} = \frac{\eta^2}{2} M_{\rm acc} c^2. 
   \label{eq:Eabs}
\end{equation}

\begin{figure*}
	\includegraphics[width=\textwidth]{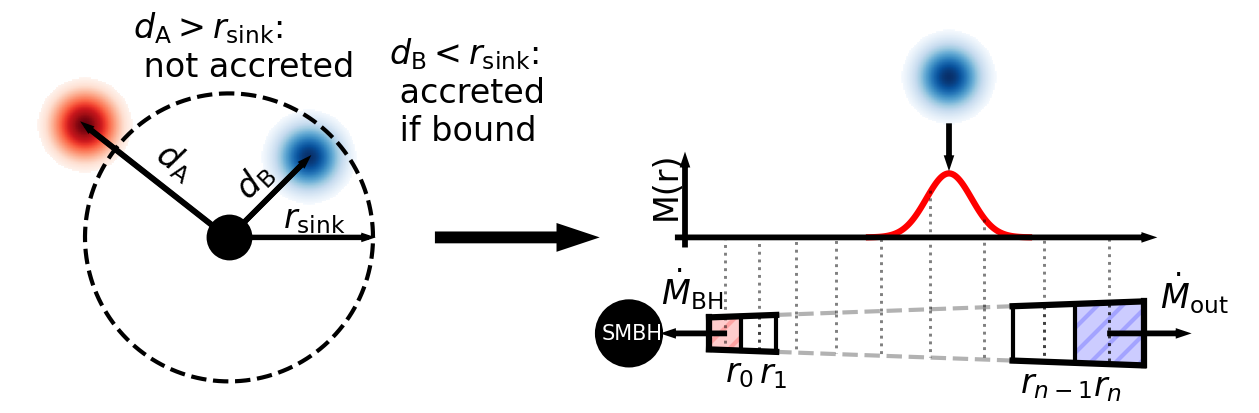}
    \caption{Visual representation of the accretion of the blue sph particle. When the particle crosses $r_{\rm sink}$ and its angular momentum is low enough it is marked for accretion. As the particle represents a gas parcel with nonzero dimensions, we distribute its mass over a number or rings of the accretion disc. The amount of matter added to each ring is weighted by the kernel used in the simulation. The red particle, conversely, is not accreted. }
    \label{fig:GeoFeed}
\end{figure*}

\section{Accretion disc particle} \label{sec:AccPartSetup}

The core of our prescription is a one-dimensional thin accretion disc model which we couple to the SMBH sink particle in the hydrodynamical simulation. Here, we present the details of this model.

The geometry of the accretion disc is defined by a number of concentric rings of logarithmically increasing radius as shown in Fig. \ref{fig:GeoFeed}. The innermost and outermost annuli are used as outflow-type boundaries. The inner lies inwards of the the innermost stable orbit and the outermost outwards of the sink radius of the particle. 
We make a simplifying assumption that the geometry of the accretion disc is fixed, that is, we do not need to adjust the radii of the annuli as the mass of the black hole grows, since the growth is negligible when compared to the initial SMBH mass. 

As is usual for a two-stage approach, the mass of particles that cross some predetermined $r_{\rm{sink}}$ and fulfil accretion criteria is added to the accretion disc. As the smoothing length of the accreted SPH particle can be comparable to the size of the disc, some care needs to be taken when adding the accreted material on to the disc. Our solution is to smoothly distribute the accreted material over a number of rings spanned by the particle, as illustrated in fig. \ref{fig:GeoFeed}. We centre the distribution defined by the kernel of the SPH particle around the circularization radius of the accreted particle given by $R_{\rm circ} = J_{\rm part}^2 / (G M_{\rm BH}$), where $J_{\rm part}$ is the angular momentum per unit mass of the accreted particle. The portion of the distribution that would end up outside the disc is put on the outer boundary. At the inner boundary, the distribution is reflected. This is a compromise between a more rigorous calculation and the simple mass injection to a single ring.

We evolve the disc using a pseudo-Newtonian Paczy\'nsky-Wiita (PW) potential \cite{Paczy1980}:
\begin{equation}
 \phi = \frac{-G M_{\rm BH}}{R-R_{\rm g}},
 \label{eq:PWpot}
 \end{equation}
where $R$ is the distance from the SMBH and $R_{\rm g}$ is the Schwarzschild radius. The equation for the viscous evolution of the disc is then:
\begin{equation}
\frac{\partial \Sigma}{\partial t} = \frac{3}{R}\frac{\partial }{\partial R} \left [ \frac{(R - R_{\rm g})^2}{R^{1/2}(R-3R_{\rm g})}\
 \frac{\partial}{\partial R} \left ( \nu \Sigma R^{3/2} \frac{R-\frac{1}{3}R_{\rm g}}{(R-R_{\rm g})^2} \right )  \right ];
 \label{eq:Diff}
\end{equation}
with viscosity defined as $\nu = \alpha c_{\rm s} H$, where $c_{\rm s}$ is the speed of sound, $H$ is the height of the disc and $\alpha=0.1$ \citep{ShakuraSunyaev} \footnote{See Appendix \ref{App:Derivations} for a detailed derivation of this and other relevant disc equations.}.

We solve the diffusion equation numerically, updating the surface density in each annulus comprising the accretion disc $\Sigma(R_i, t)$ after each timestep:
 \begin{equation}
 \Sigma(R_i, t+\Delta t) = \Sigma(R_i, t) + \frac{\partial \Sigma(R_i, t)}{\partial t} \Delta t;
 \label{eq:finiteDiff}
 \end{equation}
Accretion disc evolution occurs after each sink particle step. The dynamical timescale of the accretion disc can be much smaller than that of the simulation as a whole so we use a different criterion to determine a timestep appropriate for the disc. An initial timestep estimate is calculated by:
\begin{equation}
    \Delta t_{\rm est} = \rm{C}\, \rm{min}\left[\Delta t_i\right] = \rm{C}\, \rm{min}\left[\frac{\Delta R_i^2}{ \nu_i}\right], 
\end{equation}
with an arbitrary Courant factor, in our case $\rm{C} = 0.01$ works well. The actual $\Delta t$ is then defined in reference to the sink particle's timestep $\Delta t_{\rm sink}$:
\begin{align}
    n_{\rm steps} &= \textrm{ceil}\left[ \frac{\Delta t_{\rm sink}}{\Delta t_{\rm est}} \right] + 1, \\
    \Delta t &= \frac{\Delta t_{\rm sink} }{n_{\rm steps}},
\end{align}
to ensure that both $\Delta t n_{\rm steps} = \Delta t_{\rm sink}$ and $\Delta t < \Delta t_{\rm est}$ conditions are met. Here $\textrm{ceil}\left[x\right]$ is the ceiling function.

After a few steps, the properties of the accretion disc change and it may turn out that $\Delta t$ is too large to produce a stable result. In this case the timestep is halved and the system is restored to a configuration at the end of the sink particle step, repeating the previous steps and continuing with the corrected timestep. Note that only the viscous evolution of the accretion disc is performed using these, generally smaller, timesteps. Both the accretion on to the disc and the ejection of feedback energy-momentum packets occur at the main simulation timesteps.

\begin{table*}
\begin{tabular}{@{}LLS[table-format=1.1]S[table-format=2.1]S[table-format=3.0]S[table-format=3.0]S[table-format=1.2]S[table-format=1.2]S[table-format=1.2]S[table-format=1.2]@{}}
\toprule
  \textnormal{Run} &
  \textnormal{FB} &
  \multicolumn{1}{c}{$E_{\rm tot.o}$} &
  \multicolumn{1}{c}{$E_{\rm tot.i}$} &
  \multicolumn{1}{l}{$t_{\rm Edd}$} &
  \multicolumn{1}{r}{$t_{\rm stop}$} &
  \multicolumn{1}{l}{$M_{\rm acc.tot.}$} &
 \multicolumn{1}{l}{ $M_{\rm acc.bh.}$} &
  \multicolumn{1}{l}{$M_{\rm acc.esc.}$} &
  \multicolumn{1}{r}{$M_{\rm peak.disc.} $ }\\ \cmidrule(ll){5-6}
  \cmidrule(ll){7-10}
   &
  &
  \multicolumn{1}{l}{$10^{57}$ erg} &
 \multicolumn{1}{l}{ $10^{55}$ erg} &
  \multicolumn{2}{c}{kyr} 
   &
  \multicolumn{4}{c}{$10^5\msun{}$} \\
  \midrule
\multicolumn{1}{l}{\texttt{nFBr0}} & \textnormal{off} & 5.3 & \multicolumn{1}{c}{\text{-}} & 98  & \multicolumn{1}{c}{\text{-}}   & 1.17 & 0.44 & 0.60 & 0.42 \\
\multicolumn{1}{l}{\texttt{nFBr1}} & \textnormal{off} & 5.3 & \multicolumn{1}{c}{\text{-}} & 163 & \multicolumn{1}{c}{\text{-}}   & 1.18 & 0.45 & 0.62 & 0.41 \\
\multicolumn{1}{l}{\texttt{nFBr2}} & \textnormal{off} & 5.0 & \multicolumn{1}{c}{\text{-}} & 137 & \multicolumn{1}{c}{\text{-}}   & 1.10 & 0.41 & 0.58 & 0.46 \\
\multicolumn{1}{l}{\texttt{nFBr3}} & \textnormal{off} & 5.9 & \multicolumn{1}{c}{\text{-}} & 145 & \multicolumn{1}{c}{\text{-}}   & 1.29 & 0.48 & 0.68 & 0.46 \\
\multicolumn{1}{l}{\texttt{FBr0}} & \textnormal{on}  & 2.5 & 7.3 & 0   & 181 & 0.55 & 0.21 & 0.28 & 0.32 \\
\multicolumn{1}{l}{\texttt{FBr1}} & \textnormal{on}  & 3.2 & 9.2 & 3   & 276 & 0.69 & 0.26 & 0.35 & 0.32 \\
    \multicolumn{1}{l}{\texttt{FBr2}} & \textnormal{on}  & 3.3 & 9.7 & 66  & 207 & 0.71 & 0.28 & 0.37 & 0.39 \\
\multicolumn{1}{l}{\texttt{FBr3}} & \textnormal{on}  & 3.2 & 9.1 & 32  & 219 & 0.69 & 0.26 & 0.35 & 0.37 \\ \bottomrule
\multicolumn{1}{l}{\texttt{INSTr0}} & \textnormal{on}  & 3.3 & 15.9 & 107 & \multicolumn{1}{c}{\text{-}} & 0.44 & 0.44 & \multicolumn{1}{c}{\text{-}} & \multicolumn{1}{c}{\text{-}} \\ 
\multicolumn{1}{l}{\texttt{INSTr1}} & \textnormal{on}  & 3.4 & 16.3 & 117 & \multicolumn{1}{c}{\text{-}} & 0.63 & 0.63 & \multicolumn{1}{c}{\text{-}} & \multicolumn{1}{c}{\text{-}} \\ 
\multicolumn{1}{l}{\texttt{INSTr2}} & \textnormal{on}  & 3.5 & 16.5 & 140 & \multicolumn{1}{c}{\text{-}} & 0.68 & 0.68 & \multicolumn{1}{c}{\text{-}} & \multicolumn{1}{c}{\text{-}}\\ 
\multicolumn{1}{l}{\texttt{INSTr3}} & \textnormal{on}  & 3.5 & 16.8 & 142 & \multicolumn{1}{c}{\text{-}} & 0.63 & 0.63 & \multicolumn{1}{c}{\text{-}} & \multicolumn{1}{c}{\text{-}} \\ \bottomrule\bottomrule
\end{tabular}
\caption{ Main results of all simulations. All values calculated at $t = 500$~kyr. FB column shows whether feedback is turned on in the run. The total energy generated by the accretion disc or sink particle is $E_{\rm tot.o}$ and the total  energy absorbed by the surrounding gas is $E_{\rm tot.i}$ (note that $\eta=0.1$ for the INST runs). $t_{\rm Edd}$ is the time spent at or over the Eddington luminosity and $t_{\rm stop}$ is the time at which the central cavity appears and the disc accretion fully stops. The total accreted mass is given by $M_{\rm acc.tot.}$, total mass fed to the SMBH is $M_{\rm acc.bh.}$, the total amount of mass escaping via the outer disc boundary is $M_{\rm acc.esc.}$ and the maximum amount of matter contained in the accretion disc is $M_{\rm peak.esc.}$. }
\label{tab:ResultSummary}
\end{table*}

\begin{figure*}
    \begin{centering}
        \centering
        \begin{centering}
        \includegraphics[width=\textwidth]{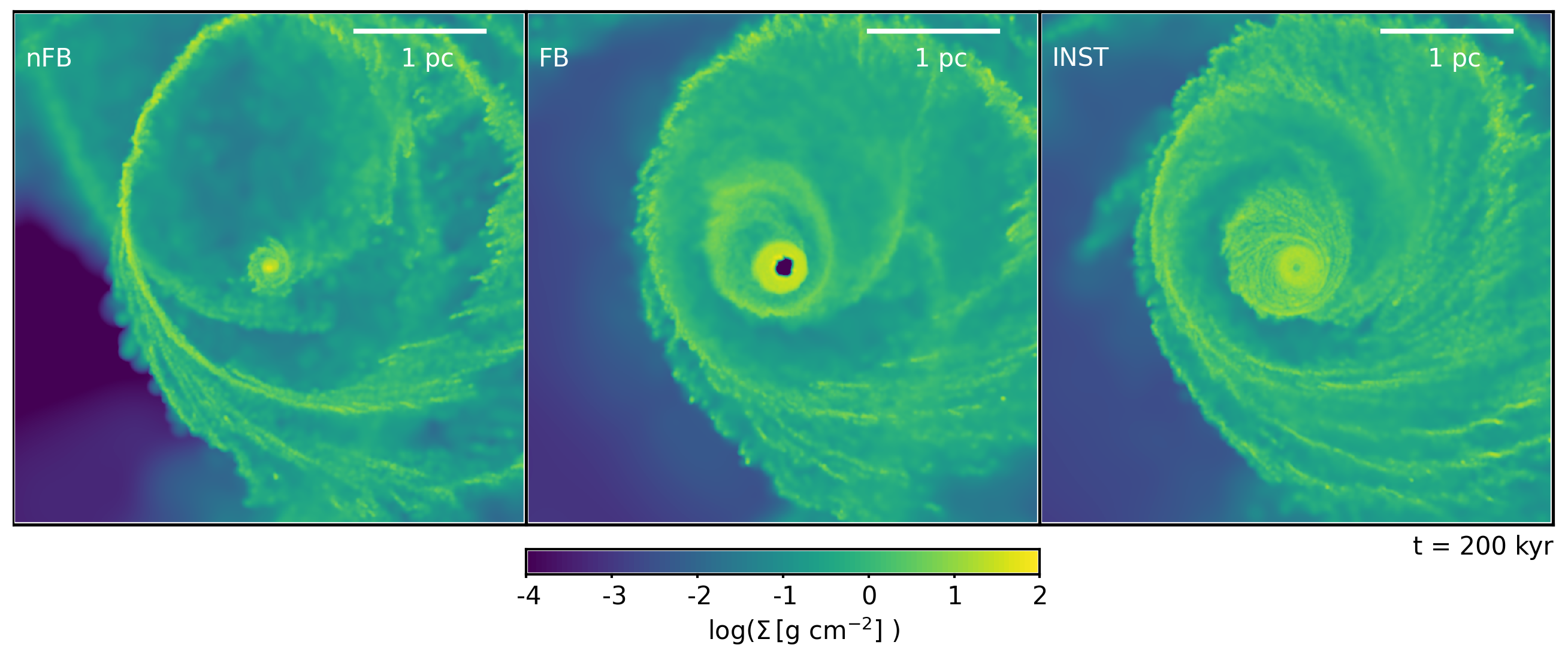}
        \end{centering}
        \caption{Density maps of the zoomed in $4$~pc region in \texttt{nFBr0}, \texttt{FBr0} and \texttt{INSTr0} runs at time $t=200$~kyr. Clear differences are already apparent - a well defined central disc aligned  to the $xy$-plane in simulations with feedback is in contrast to the more compact and misaligned central in the nFB run. The central cavity is also already present in the FB run.}
        \label{fig:3Diff} 
        \end{centering}
\end{figure*}

After each disc timestep, the parameters of the accretion disc are also updated. The PW potential yields the following expression for the viscous dissipation per unit disc face area $D(R)$:
\begin{equation}
    D(R) = \frac{3}{8} \frac{\dot{M}}{\pi} \frac{G M_{\rm BH} }{ R}  \left( \frac{1}{R-R_{\rm g}} - \frac{3^{3/2} R_{\rm g}^{1/2}}{2 R^{3/2}} \right)  \frac{\left( R-R_{\rm g} \right )^2}{ R-\frac{1}{3}R_{\rm g}},
\end{equation}
which we use to calculate the luminosity of each annulus:
\begin{equation}
 \label{eq:AccDiscLum}
    L(R) = 2 \pi \left[D\left(R_{\rm out}\right)R_{\rm out} + D\left(R_{\rm in}\right)R_{\rm in}\right] \left(R_{\rm out} - R_{\rm in}\right) 
\end{equation}
here $R_{\rm in}$ and $R_{\rm out}$ are, respectively, the inner and outer radii of the $i$-th annulus. Summing over all annuli gives the total luminosity of the accretion disc which we use to determine the amount of energy and/or momentum to be injected into the surrounding gas. The radiative efficiency of accretion stays close to the expected value for the PW potential $\eta \sim 6.25\%$ \citep{Paczy1980} throughout our simulation runs. This is somewhat higher than $\eta = 5.7\%$ appropriate for a Schwarzschild black hole; on the other hand, any real black hole is expected to be at least somewhat rotating, which increases the radiative efficiency.

\section{Results }\label{results}

We ran 12 simulations in total: four with (FB) and four without feedback (nFB) using our accretion disc method, plus four runs with feedback using an instantaneous accretion prescription (INST). In the latter group, we set the radiative efficiency to the usual value $\eta=0.1$, calculate the luminosity using $L = \eta \dot{M}_{\rm BH} c^2$ and limit it by the Eddington limit. Note that since the matter fed to the disc is assumed to be instantaneously transported to the SMBH in the INST case, the nFB disc feeding rate would correspond to the SMBH accretion rate in an INST run without feedback. This makes the additional four runs of INST without feedback redundant. The differences among the four simulations in each group are only stochastic: the cloud and ring are realised from different initial distributions of particles. We performed these stochastically differing simulations in order to understand which of our results are robust and represent actual differences in the accretion methods. The key results are summarized in Table \ref{tab:ResultSummary}.

All runs produce qualitatively similar behaviour, with a single clear outlier \texttt{FBr2}. In general, the system evolves as follows \footnote{Density maps representative of the general evolution are presented in Appendix \ref{app:grids}}. It takes about $40$~kyr for the cloud to reach the torus. Due to the initial turbulence seeding, at this time there are already clumps and filaments present and some of portion of the cloud misses the ring; this material returns to the system later via an elongated stream. The cloud impacts the torus in the direction opposite to its rotation, thus a significant portion of gas loses angular momentum and is quickly transported to the accretion disc, with the disc accretion rate exceeding the Eddington mass accretion rate of the SMBH by almost an order of magnitude. Later on, a central disc/ring system forms within approximately the central parsec, surrounded by a larger ring that is still in the process of settling by $t = 500$~kyr. However, some identifiable differences between simulation sets begin to appear after a few tens of kyr following the initial collision. In Fig. \ref{fig:3Diff} we compare a representative run from each set at $t=200$~kyr; around this time a central cavity becomes clearly apparent in the FB runs. At this point the nFB run has a more compact ($r_{\rm out}\sim0.1$~pc) central structure which is slightly warped in the centre. In contrast, both FB and INST produce a more extended central structure ($r_{\rm out} \gtrsim 0.2$~pc), with a central cavity appearing only in FB runs. Over time the differences in the extent of the central structure seem to mostly disappear (eg. Fig. (\ref{fig:grid_00}-\ref{fig:grid_00i}) but the nFB central discs remain more warped and central cavities in the FB runs become even more pronounced.

\begin{figure*}
	\includegraphics[width=\textwidth]{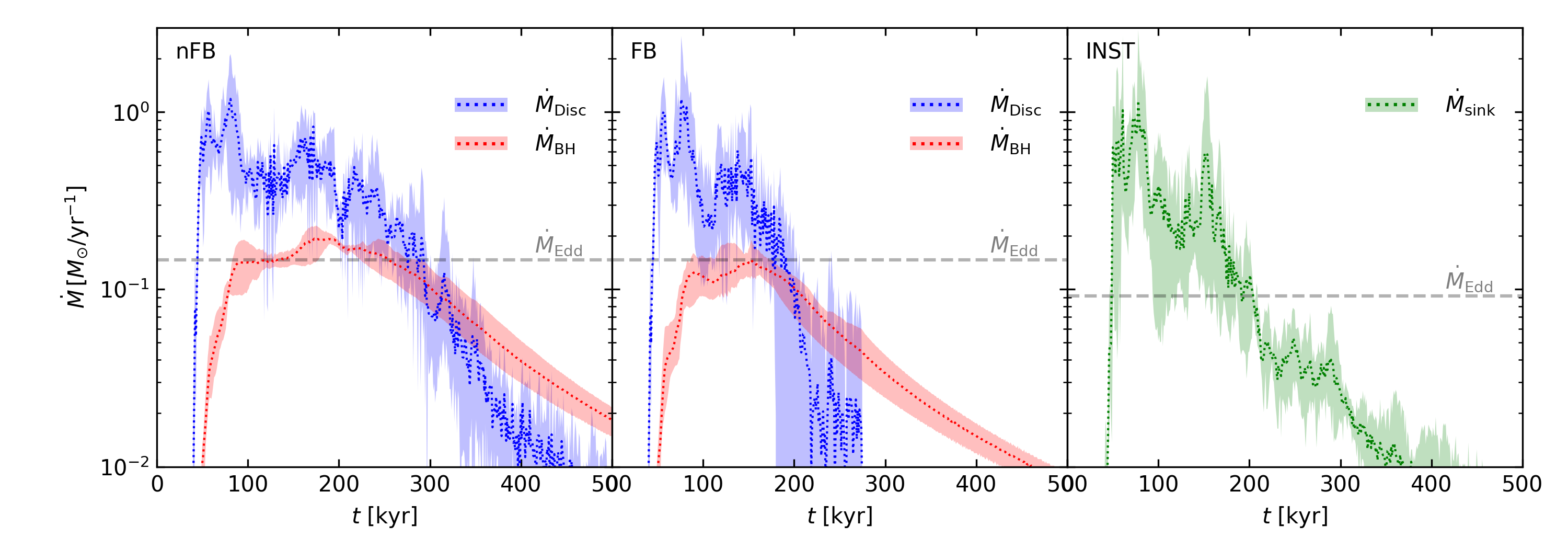}
	\vspace{-0.75cm}
    \caption{Accretion disc (blue) and SMBH (red) accretion rates in nFB (left) and FB (middle) runs and sink accretion rate in the INST runs (right). The dotted lines are the average rates in each set, while the filled regions show the effect of stochastic variations in each set. The grey horizontal dashed line shows the Eddington mass accretion rate $\dot{M}_{\rm Edd}=L_{\rm Edd} \eta^{-1}  c^{-2}$}
    \label{fig:AccRates_var}
\end{figure*}

\begin{figure}
	\includegraphics[width=\columnwidth]{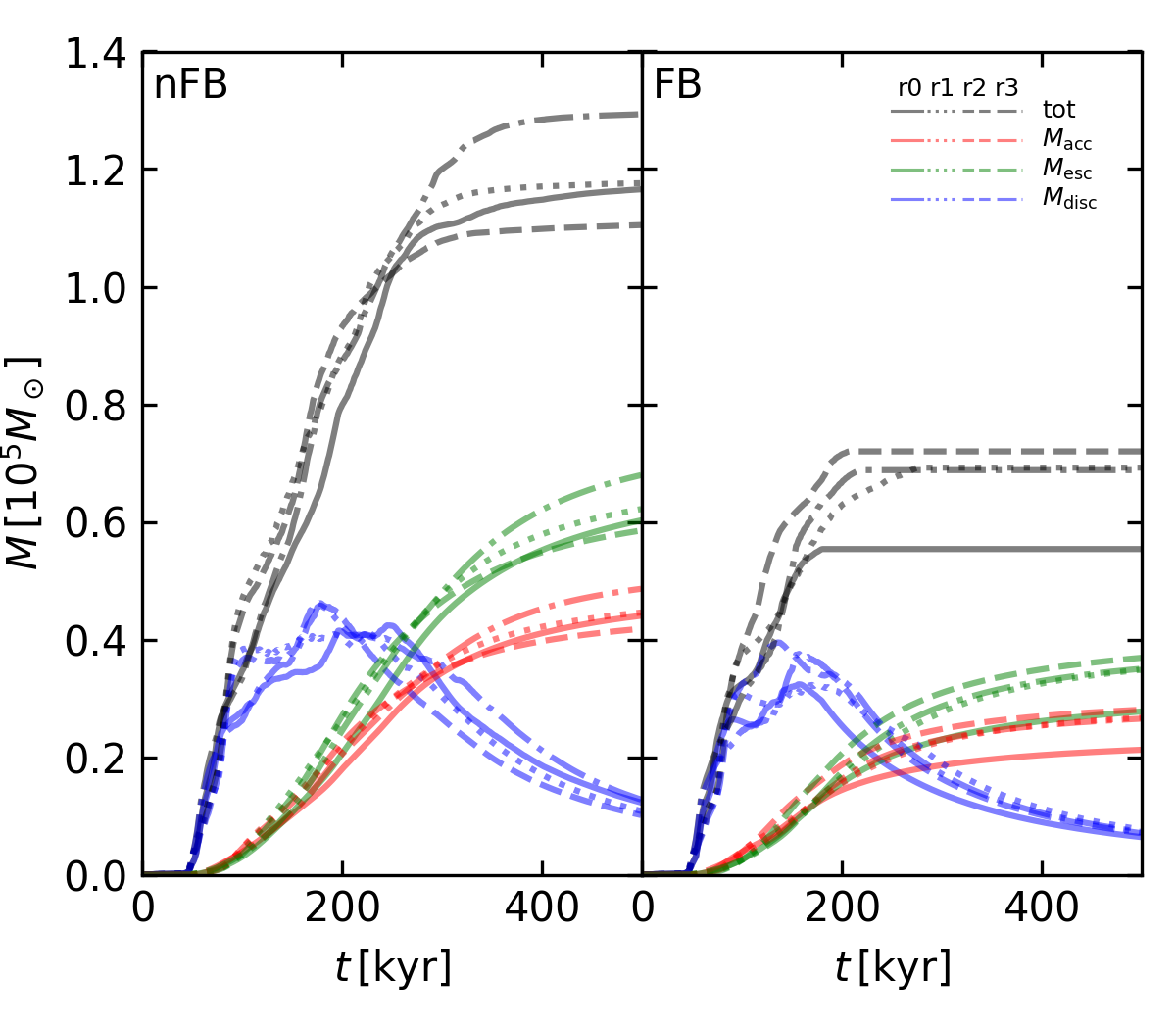}
	\vspace{-0.75cm}
    \caption{Mass growth curves of mass accreted by the accretion disc particle (black), mass contained within the accretion disc (blue), mass added to the SMBH (red) and mass that escapes the disc via outer boundary (green). The left-hand panel shows results for the nFB simulations while the right-hand panel shows results of FB simulations. Different line styles represent different realizations.}
    \label{fig:MassGrowth}
\end{figure}

A comparison between accretion rates in all runs is shown in Fig. \ref{fig:AccRates_var}. The blue in nFB and FB runs corresponds to the accretion disc feeding rate, while red shows the SMBH feeding rate; in the INST case, green shows the feeding rate on to the SMBH sink particle. In all cases the dotted lines are mean values of the four simulations in the set, and the filled regions encompass the range of variations in the four simulations. We can see that accretion on to the disc/sink particle is relatively highly variable with differences by a factor two or more common on timescales $<10$~kyr, which is due to the chaotic nature of clumpy matter infall. In contrast, the disc-mediated feeding of the SMBH is relatively smooth and slightly delayed in time. In FB runs, the feeding of the disc comes to an abrupt stop between $\sim 180-280$~kyr, which can be tied to the appearance of central cavities, while in the nFB runs the disc particle is still being fed after this time at a gradually decreasing rate. We examine the nFB and FB runs further in Fig. \ref{fig:MassGrowth} where we show the amount of matter contained within each reservoir of our sink particle. Here, the total amount of gas fed to the disc is shown in black, while the amount of matter feeding the SMBH and escaping via the outer boundary is shown in red and green respectively. We can see the disc itself (blue lines) is far from empty at the end of the rapid accretion stage, so the feeding of the SMBH continues for $\sim 200$~kyr more even after the central region is cleared of gas. The feeding of the SMBH sink particle continues also in the INST runs, which did not produce the central cavity, but the rate of accretion is strongly diminished, generally to lower values than the SMBH accretion rate in the FB runs.

In total, between $\sim 1.1\times10^5 \msun{}$ and $\sim 1.3\times10^5\msun{}$ feeds the disc particle in the nFB runs; this is a significant fraction of the initial gas mass ($\sim 2\times10^5\msun{}$). More than a third of this mass ends up feeding the SMBH and about a half escapes the disc via the outer boundary (Fig. \ref{fig:MassGrowth}). Similar proportions are found in the FB runs, with the total amount of accretion reduced by about a half. The commonality between these proportions in all runs suggests that the stochasticity and the relatively chaotic dynamics of the surrounding gas have little effect on the accretion disc itself. We explore how our choice of minimum smoothing length effect the disc(sec. \cref{sec:Param_tune}) and how the escaping gas might affect the system (sec. \cref{sec:lost_gas}) in the Discussion.

\begin{figure}
	\includegraphics[width=\columnwidth]{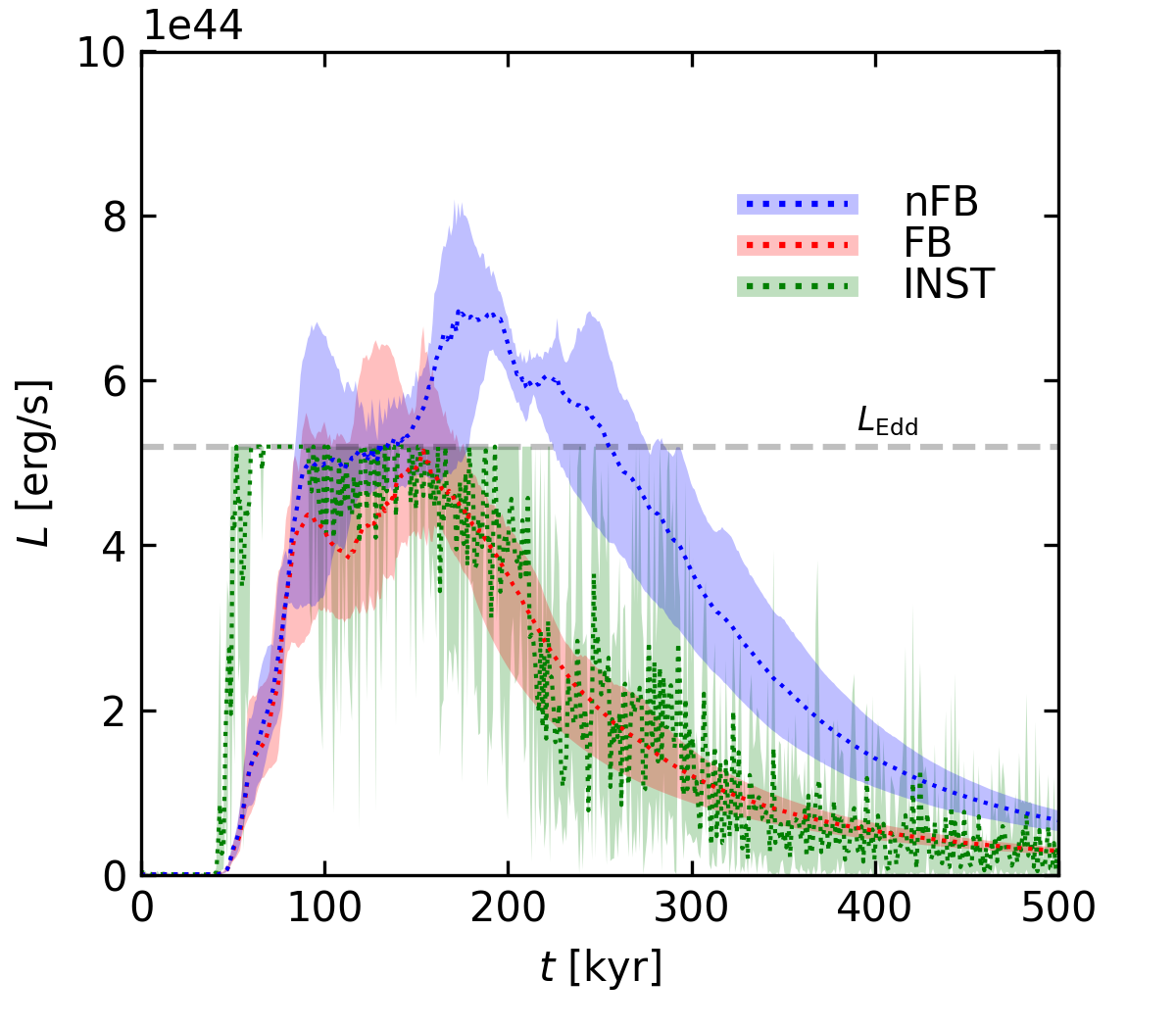}
	\vspace{-0.75cm}
    \caption{ Accretion disc luminosity change over time. Dotted curves show the mean in a set of of runs, while the filled regions show the variation found between runs in a set. Eddington limit is shown by the gray dashed line.}
    \label{fig:Luminosity_var}
\end{figure}

The evolution of luminosity in the three simulation sets is shown in Fig. (\ref{fig:Luminosity_var}). The stalled accretion also results in a similarly decreased luminosity in both sets with feedback when compared to the nFB runs\footnote{Note that in the INST simulations, we compute the luminosity of the accretion disc, but do not generate feedback particles, thus it has no effect on the surrounding gas.}. Luminosity in INST simulations is highly variable as it follows the variations of accretion on to the sink particle. Conversely, the evolution of disc luminosity in both FB and nFB is relatively smooth, since it follows the feeding of the SMBH; the radiative efficiency stays nearly constant at $\eta\approx6.3$~\% during the period of activity. Accretion disc luminosities reach super-Eddington values in all except one FB run. While feedback significantly reduces gas infall, thus the super-Eddington phases are not sustained for long and the peak luminosity is $< 30\%$ (60\% in nFB) higher than  $\le{}$ , we note that our simple model does include any mass loss from the disc itself that occurs in the super-Eddington regime. More tests are required to determine if some artificial limit on luminosity in our sub-resolution prescription is warranted.

\begin{figure*}
    \begin{centering}
        \centering
        \begin{centering}
        \includegraphics[width=\textwidth]{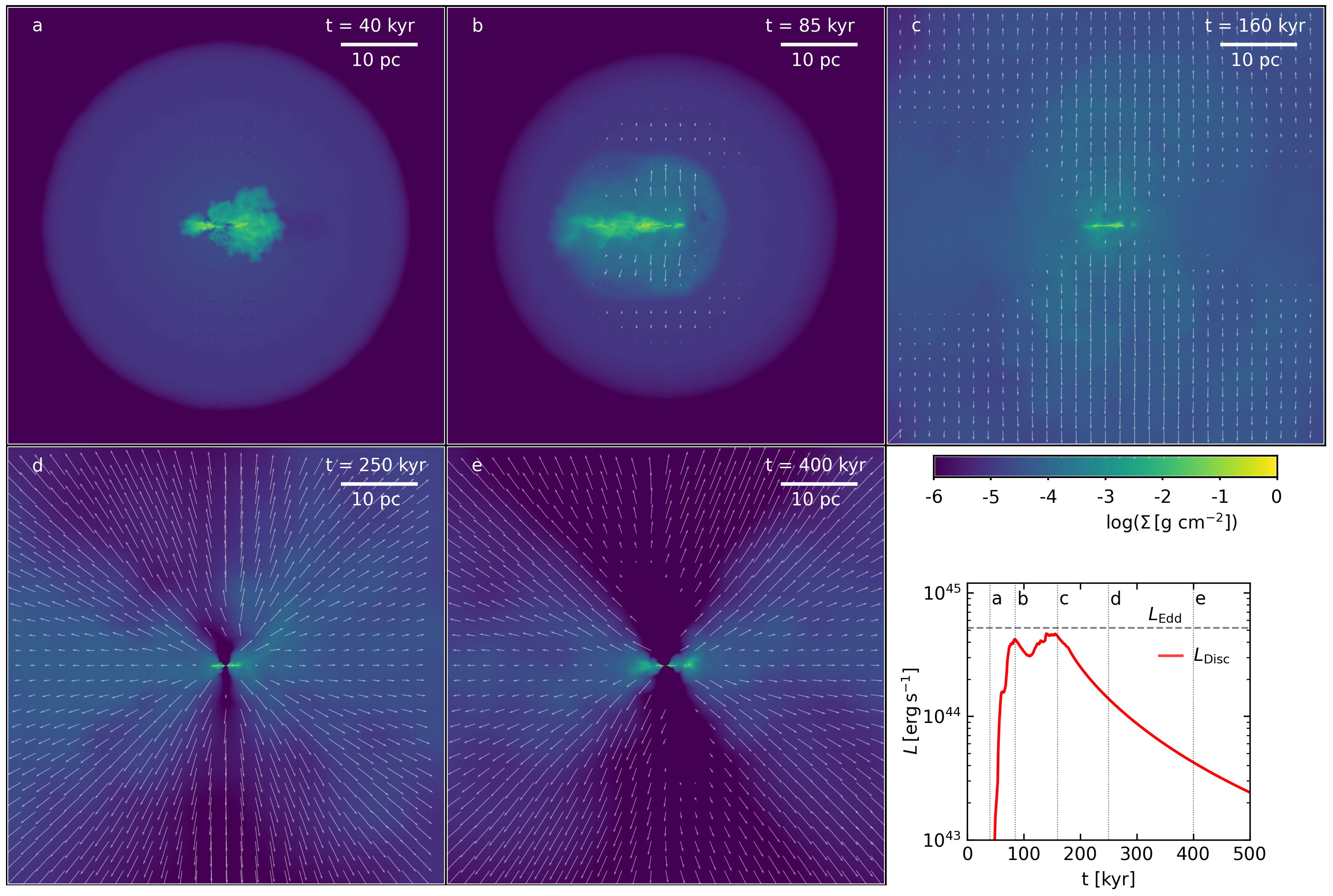}
        \end{centering}
        \caption{Surface density maps at various times during the run \texttt{FBr0}. The view is from the side of the initial ring plane. The change in luminosity over time is shown at the bottom right, along with the droplines corresponding to the times of the density map. Vectors show direction of gas movement, their length proportional to logarithm of gas velocity}
        \label{fig:outflow} 
        \end{centering}
\end{figure*}

The disc in the centre of the system has a collimating effect on the isotropically emitted SMBH wind; this leads to a conical outflow (Fig. \ref{fig:outflow}). The total energy produced by the accretion disc in the FB runs is between $2.5\times10^{57}$~erg and  $3\times10^{57}$~erg, but only a portion of this energy is absorbed by the surrounding ISM in accordance with (\ref{eq:Eabs}), so the actual energy injected in gas particles in our simulation is about $7.8\times10^{55}$~erg to  $10.3\times10^{55}$~erg. This value is still a few orders of magnitude higher than the estimate for progenitor event of the 430-pc radio bubbles (\citep[$7\times10^{52}$~erg]{430RADIOBUBBLES}). It also exceeds the energy required to form the Fermi Bubbles by at least a factor of a few \citep[$\sim10^{54-55}$~erg]{fermior}. In all cases except one, SMBH winds are not strong enough to completely disrupt the system, leaving an intact central disc and a surrounding ring system, just with a pronounced central cavity. As the system evolves, more and more of the background gas is removed and the opening angle of the cone increases. The same general behaviour describes the INST runs, but they produce approximately twice the amount of energy - this is largely due to the larger radiative efficiency value $\eta_{\rm INST}/ \eta_{\rm FB} \simeq 1.6$. Also, as there is less of a delay between accretion and luminosity peaks in INST runs, the effect of feedback is near immediate. But, as in the case with the central cavity, there is less of a contrast between the outflow cone and the background; this suggests that in the INST runs, AGN feedback is less efficient in pushing the gas out.

\begin{figure}
	\includegraphics[width=\columnwidth]{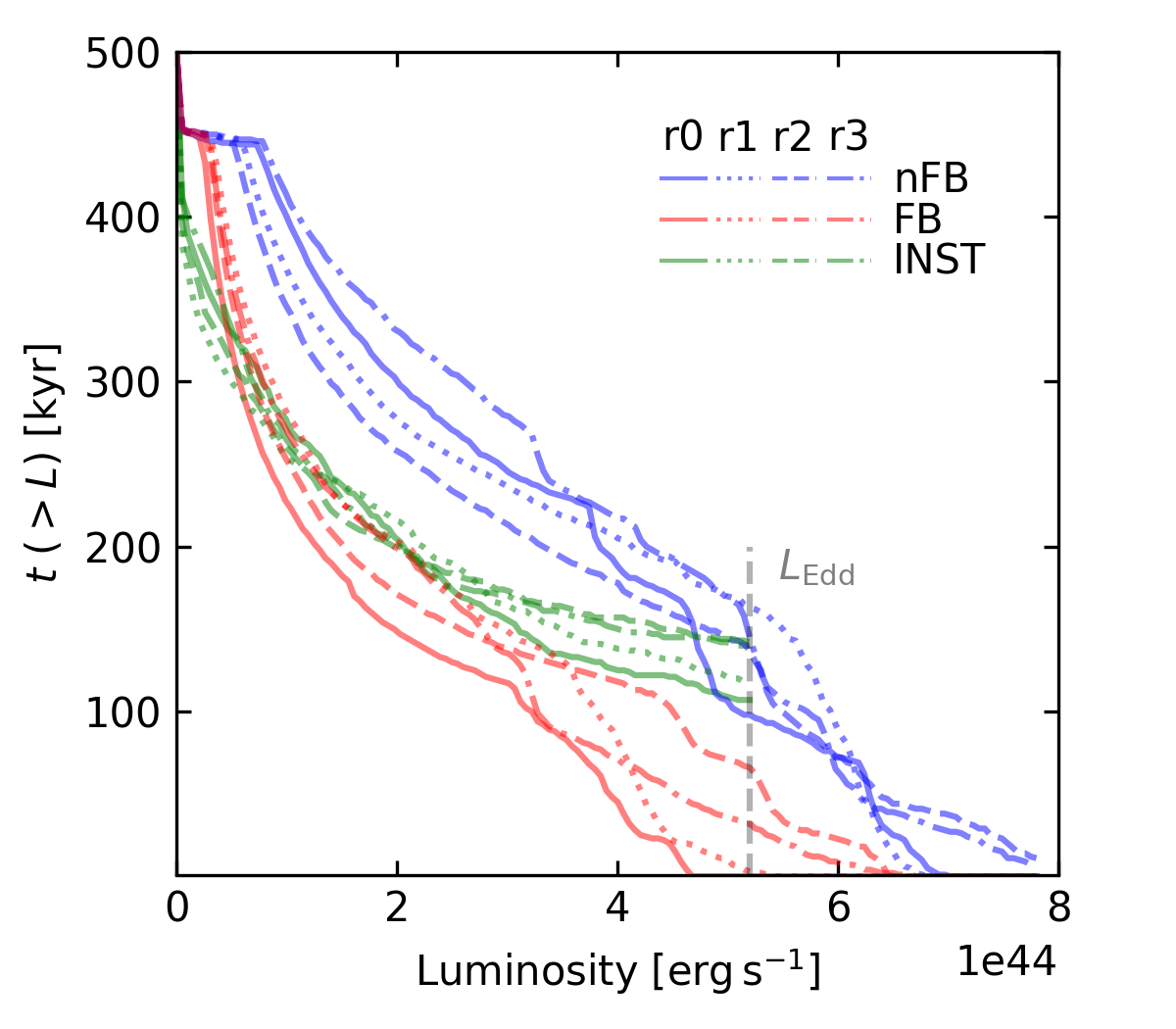}
	\vspace{-0.75cm}
    \caption{AGN luminosity functions in nFB (blue), FB (red) and INST (green) simulations. The Eddington limit is shown by the vertical gray dashed line.}
    \label{fig:time_over}
\end{figure}

By $t=500$~kyr the outer portion of the central disc and the surrounding inner filaments/rings are aligned with the $xy$-plane within a few degrees in all runs, while the inner portion of the central disc is somewhat warped - the disc rotational plane shifts up to $50^{\deg}$ in nFB and up to $20^{\deg}$ in INST. A transient misaligned ($\sim20^{\deg}$) disc is present in two FB runs. It is possible that the SMBH wind pressure on the central disc helps to keep it within the same rotational plane. In the run \texttt{FBr2} feedback pushes out all the gas (Fig. \ref{fig:grid_12}) except for a part of the inner disc. Interestingly, the total energy ejected as wind (table \ref{tab:ResultSummary}) is similar to that in the other runs, but a longer period of time was spent in the super-Eddington regime as shown by the luminosity functions of all simulations in Fig. (\ref{fig:time_over}). Overall, at least for luminosities $L>0.5 L_{\rm Edd}$, the variation of the initial particle distribution is less important in determining the luminosity function than the differences between the accretion and/or feedback prescriptions. Another interesting aspect is that the total energy radiated as feedback is considerably larger in the INST simulations; curiously, this does not result in a complete disruption of the initial system as it does in \texttt{FBr2} or even in the creation of central cavities. It appears that sporadic bursty feedback is far less efficient at ejecting the gas and stopping further SMBH accretion than continuous energy injection, even at a milder rate. 

\subsection{A comment on performance}

An important aspect of any numerical method is its computational cost. Although we did not do rigorous benchmarking, we compared the time that our different simulations take to run. On average it takes about 160 wall-clock hours to calculate each nFB run and about 140 wall-clock hours to calculate each FB run up to $t=500$~kyr on 32 CPUs (for comparison, INST runs took slightly longer, 148 wall-clock hours, on the same system and setup). In both cases the time spent on accretion disc calculations is almost negligible: on average, the fraction of time spent on tasks directly related to the disc is $\sim0.6\%$. Interestingly, the significant amount of time spent on calculating the interactions between feedback packets and gas particles in the FB runs ($\sim32\%$ on average) is more than offset by the time saved when a large number particles that would otherwise require very small time-steps are pushed out of the centre, resulting in quicker calculations.

\section{Discussion }\label{Discussion}

\subsection{Chaotic nature of model evolution}

All our simulations produce qualitatively similar results, especially as far as the large-scale morphology of the system is concerned. The stochastic variations between runs seem to mostly affect the smaller-scale central features, e.g. the central cavity, the precise shape and size of the central disc/ring structures, etc. However, a more detailed analysis reveals more evidence of chaotic behavior. For example, the \texttt{FBr2} run shows very different morphology when compared to all other FB runs, characterized by an almost complete absence of gas in the central $5$ parsecs of the system by the end of the AGN phase. The case is especially intriguing as there are no significant differences in the total accreted mass or total injected energy, which are both only $\sim3\%$ larger than in other FB runs (Table \ref{tab:ResultSummary}). The main difference seems to be that the \texttt{FBr2} feeds the disc at a relatively consistent rate for over $100$~kyr. This can be seen in Fig. \ref{fig:MassGrowth} - the black curves show the total mass injected into the disc. \texttt{FBr0} and \texttt{FBr1} runs both have a significant dip in disc accretion rate at $\sim100$~kyr while \texttt{FBr3} somewhat lags behind the other runs. The consistent rate of feeding allows \texttt{FBr2} to maintain $L_{\rm disc} \geq \le{}$ for a longer period of time (Table \ref{tab:ResultSummary}). While not as extreme as the case of \texttt{FBr2}, the run \texttt{FBr3} also has significantly less gas remaining in the central $5$~pc at $t=500$~kyr when compared with the remaining two FB runs. \texttt{FBr3} is slightly less energetic than two of its counterparts but sustains a super-Eddington luminosity for $\sim33$~kyr. This suggests that even a brief period of super-Eddington feedback is critical in the shaping of resultant system's morphology. The results of INST simulations support this argument - there we put an artificial cap at $L_{\rm AGN} = \le{}$ and no run resulted in significant removal of gas from the centre. A counterexample to this argument can be seen when comparing the runs \texttt{FBr0} and \texttt{FBr1} - even though  \texttt{FBr1} has a higher luminosity for the majority of the activity phase and the luminosity in \texttt{FBr0} never exceeds the Eddington limit, the central cavity in \texttt{FBr0} is more pronounced.

In an attempt to understand the detailed reasons behind these differences, we check whether the more continuous high luminosity keeps the surrounding gas hot making it easier for feedback to push it out. Unfortunately, the results are inconclusive. There are no clear temperature differences between the four FB runs, where the majority of gas in the inner parsec stays at about $10^4$~K throughout the simulation. Meanwhile, in both nFB and INST runs, after the peak of the AGN phase, the temperature drops significantly down to $\gtrsim10^1$~K. Nonetheless, the inner $0.1$~pc remains continuously populated by gas with temperatures over $10^4$~K in INST runs, while all FB runs had this population pushed out at the time of the appearance of the central cavity. To summarize, there are no clear trends in gas temperature that explain the difference in kinematics.

Additionally, we measure the weight of the disc and compare it with the outward force of the AGN wind pressure in each simulation with feedback \citep{WEIGHTS}. The weight of the disc is given by:
\begin{equation}
    W_{\rm disc} \sim g(R)M_{\rm disc}(<R),
\end{equation}
where $M_{\rm disc}(<R)$ is the mass contained within $R$ and $g(R)$ is the 
$g(R)$ is the gravitational acceleration at $R$: 
\begin{equation}
g\left(R\right) = \frac{\textrm{G}M_{\rm{bh}}}{R^2} + \frac{2\sigma^2}{R}.
\label{gacc}
\end{equation}
The outward force is:
\begin{equation}
    F_{\rm out} \sim \frac{\alpha(R)}{4\pi}  \frac{L_{\rm disc}}{c},
\end{equation}
here the factor $(\alpha(R))/ 4\pi$ is the ratio between the solid angle subtended by the disc from the point of view of the SMBH and the total solid angle of a sphere, which gives the fraction of AGN wind flux absorbed by the disc. We estimate $\alpha(R)$ using a wedge cut of the density field of the central disc. We take the disc's height to be the full width at half maximum in the density field perpendicular to the disc's radius. Again, we see no significant differences between FB runs - the outward force acting on the disc always outgrows the disc's weight, initially at the centre, and gradually pushes out the gas, producing the central cavity. Conversely, in INST runs the outward force remains smaller than the weight of the disc, explaining the lack of the central cavity; the artificially imposed Eddington limit is probably responsible for this.

It is difficult determine what precisely causes \texttt{FBr2} to behave so differently. The results still suggest that the morphology of the resultant system may be very sensitive to the interaction between the infalling gas and the SMBH feedback. This implies that it might be dangerous to rely on a preset and constant $t_{\rm visc}$ parameter as a slightly different choice might result in a completely different evolution of the system. This also hints that the problem of dynamical perturbations in the central few parsecs of galaxies is truly chaotic - relatively small stochastic differences may result in drastically different outcomes. A more comprehensive study varying feedback parameters and prescriptions is required to determine if this result is not an aberration.

\subsection{Parameter Tuning}\label{sec:Param_tune}

One of the main advantages of our approach is that it gives consistent results that are less reliant on free parameters, but this comes at the cost of simplicity and a small increase in computational cost. For example, we do not define a specific viscous timescale $t_{\rm visc}$ or select a specific radius of gas accretion (although we do need to specify the outer radius of the disc). The choice of these parameters can sometimes in a large part determine the outcome of a simulation as even a small increase in luminosity may result in significant changes in chaotic systems. 

\begin{figure}
    \begin{centering}
        \centering
        \begin{centering}
        \includegraphics[width=\columnwidth]{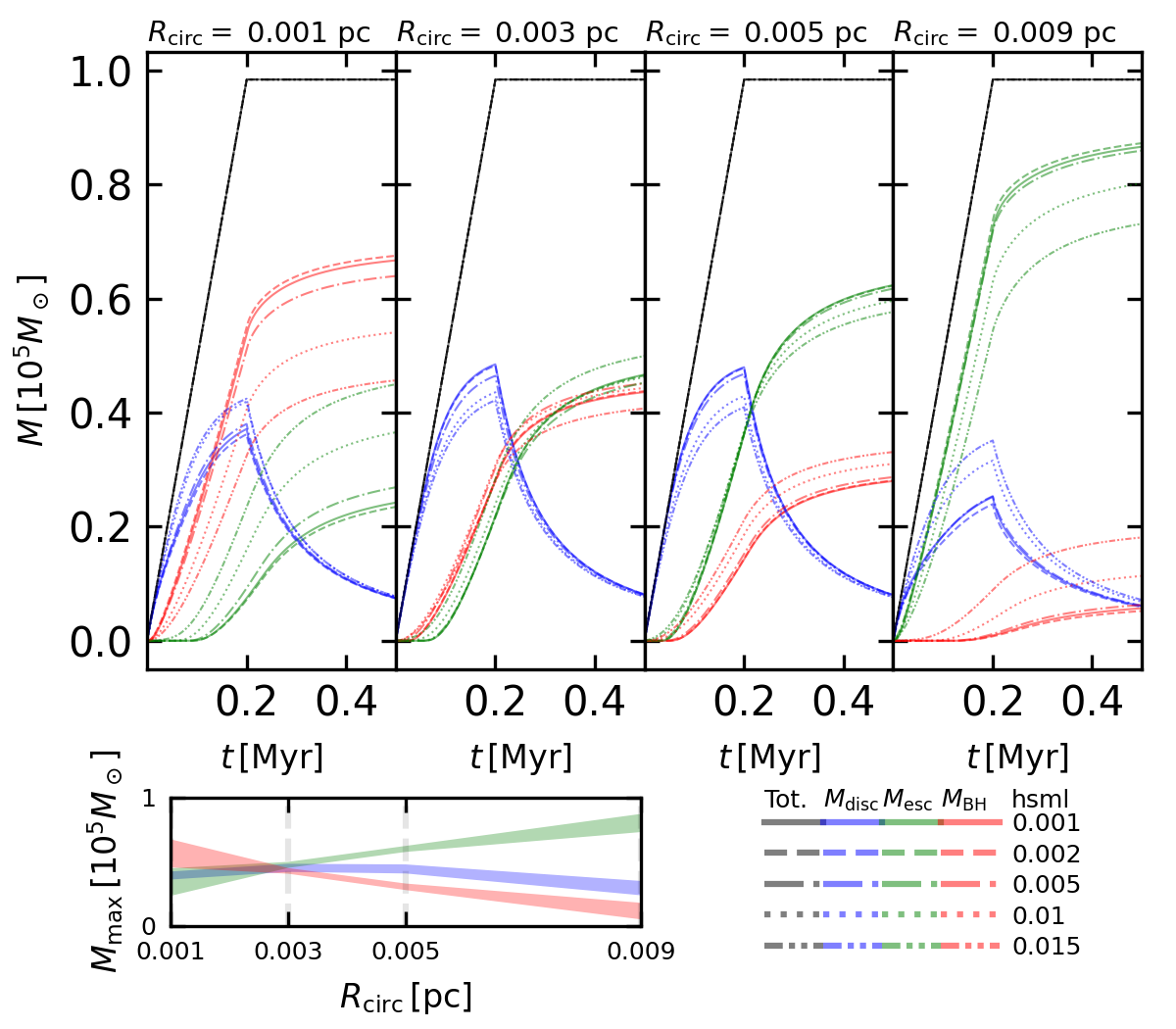}
        \end{centering}
        \caption{\label{fig:hsml_rcirc} The injected mass (black), mass in the disc (blue), mass fed to the SMBH (red) and mass escaping the disc (green) over time, depending on the injection radius ($R_{\rm circ}$, values given at the top of each of the four main panels) and smoothing length $h$ (different line styles, values given at the bottom right). The small graph on the bottom shows the peak disc mass (blue) and total mass accreted by the SMBH (red) and escaping through the outer boundary (green), with the width of the line corresponding to the variation due to the different chosen values of $h$.}
        \end{centering}
\end{figure}

In our accretion disc model, the disc evolution is set to tend towards the standard quasi-steady thin $\alpha$-accretion disc, but at each step some mass distribution is added on to the disc, depending on the properties of the accreted gas particles. Given this, it is necessary to understand how the evolution of the disc depends on this perturbation, where the distribution is determined by the parameters of the infalling matter. The angular momentum of the accreted particle is used to determine the circularization radius $R_{\rm circ}$, i.e. the radius at which its contribution is centered, and the numerical smoothing length parameter $h$ is chosen to represent the spatial extent of a particle's contribution, similar to its role in the SPH formulation of hydrodynamics. We performed a set of disc simulations with different fixed values of $R_{\rm circ}$ and $h$ \footnote{An implementation of this in \texttt{Python} is freely available in  \href{https://github.com/Caradryan/accretiondisc}{\texttt{https://github.com/Caradryan/accretiondisc}}} . The rate of disc feeding is similar to that seen in the nFB set of simulations, with $\sim10^5\msun{}$ of gas added to the accretion disc over a period of $200$~kyr at a constant rate. The resulting growth curves of the accretion disc (blue), the black hole (red) and the escaping matter (green) are shown in Fig. \ref{fig:hsml_rcirc}. Each of the four subplots at the top shows results of simulations with a different fixed $R_{\rm circ}$ value, while different line-styles correspond to different values of $h$. Here we clearly see several trends, which are all quite intuitive. First, the closer to the center a particle is fed to the disc, the more material ends up feeding the SMBH, and conversely, the further out the particle is inserted, the more mass escapes the disc via the outer boundary. The smoothing parameter $h$ in this case acts to soften this effect, as higher values allow for more matter to be placed in the middle annuli of the disc (Note the dash-double-dot line in Fig. \ref{fig:hsml_rcirc}). Interestingly, the nFB set of SPH simulations show results somewhat similar to those of steady accretion between $R_{\rm circ}=0.003$~pc and $R_{\rm circ}=0.005$~pc (Fig. \ref{fig:MassGrowth}) meaning that few accreted particles had very small or very large angular momenta. This being the case, we can also infer from Fig. \ref{fig:hsml_rcirc} that our chosen value for the minimum smoothing length $h_{\rm min} = 0.01$~pc had limited effect on the results: while some of the matter was put directly into the SMBH and the outer boundary, the amount was not overwhelming. In fact, even a significantly larger value $h = 0.015$~pc in the idealised tests did not have an impact larger than the stochastic variability between SPH simulations.

\begin{figure}
    \begin{centering}
        \centering
        \begin{centering}
        \includegraphics[width=\columnwidth]{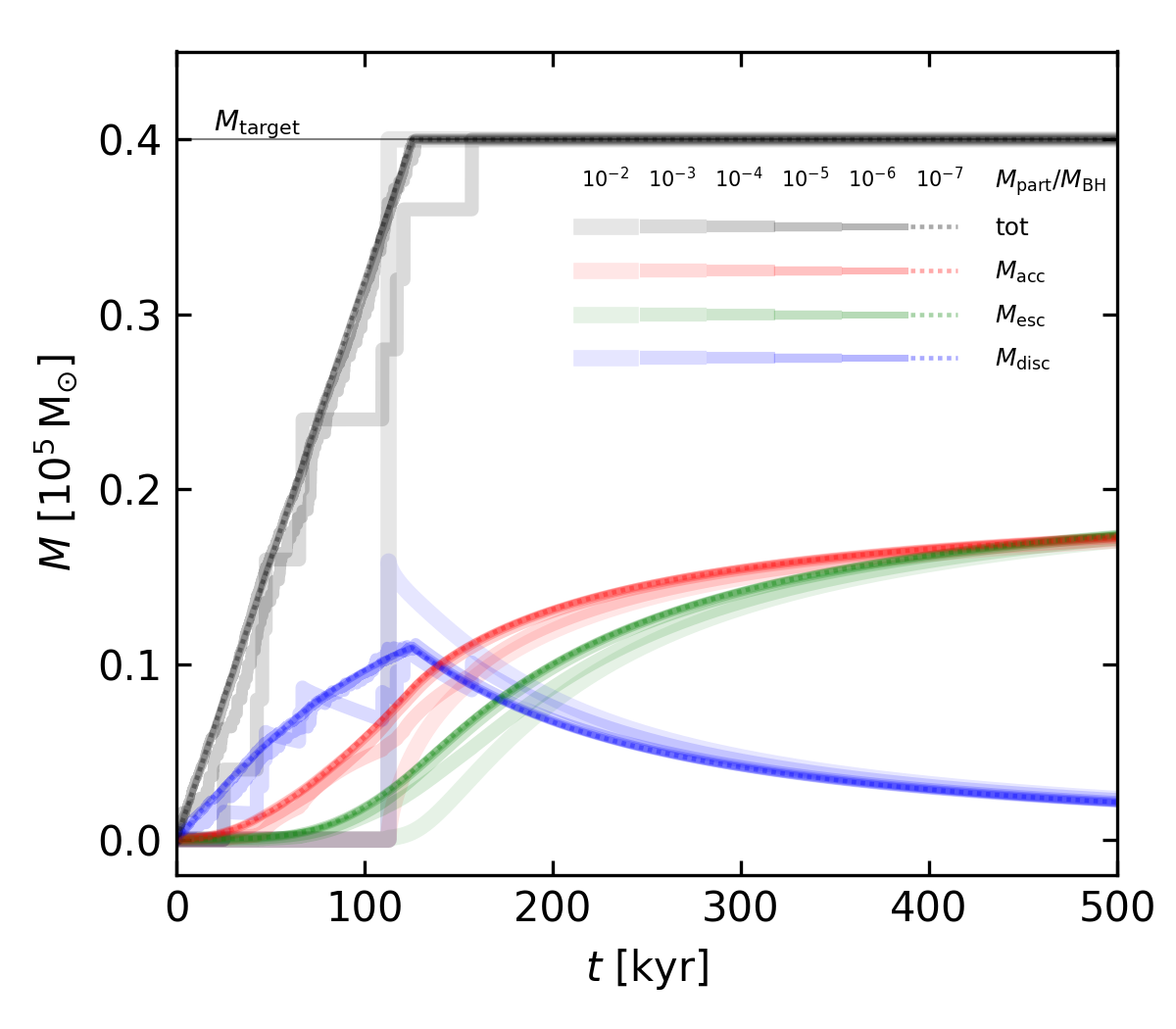}
        \end{centering}
        \caption{\label{fig:MassGrowth_mpart_diff} The injected mass (black), mass in the disc (blue), mass fed to the SMBH (red) and mass escaping the disc (green) over time, depending on the portion size of the injected matter.}
        \end{centering}
\end{figure}

\begin{figure}
    \begin{centering}
        \centering
        \begin{centering}
        \includegraphics[width=\columnwidth]{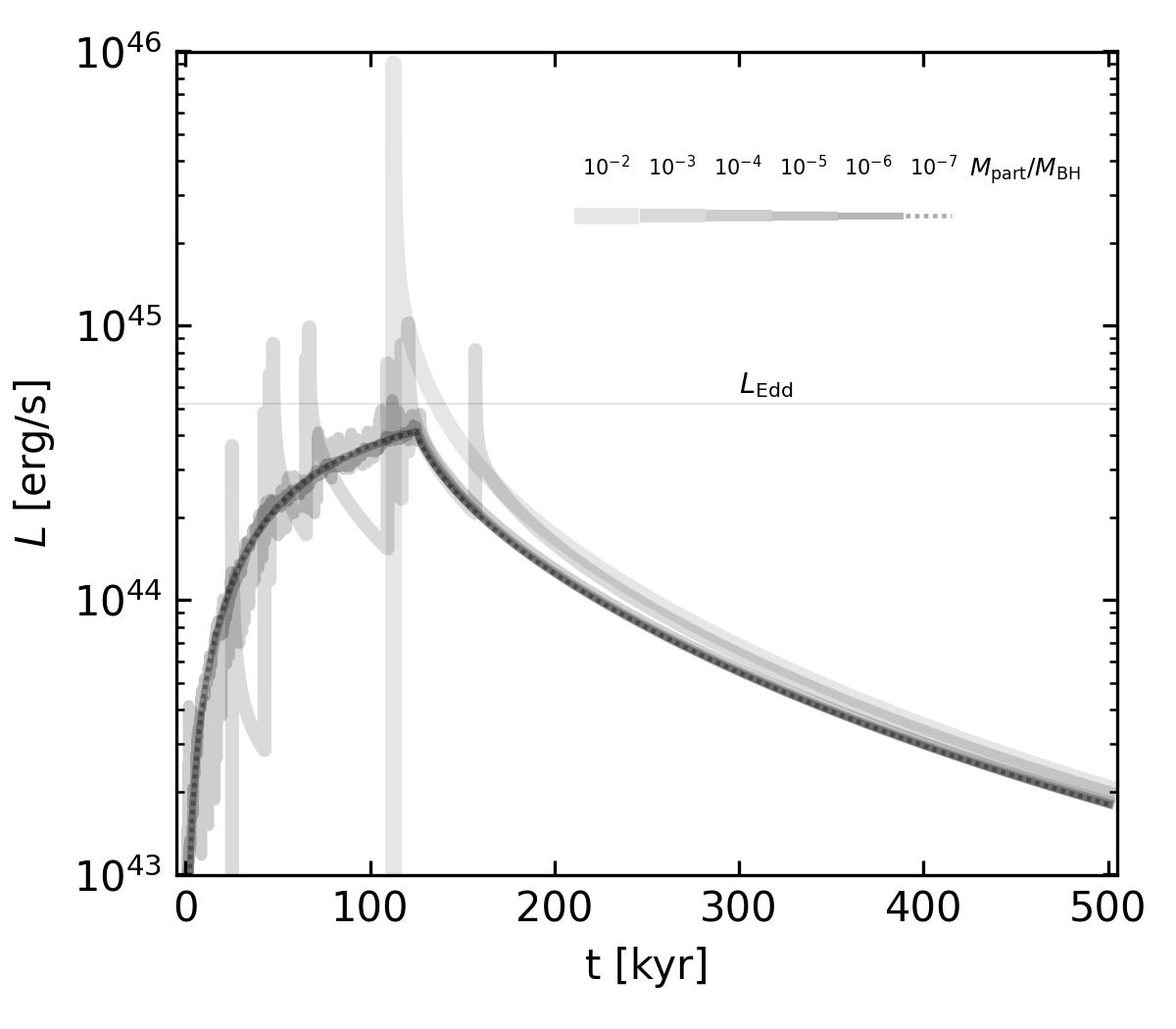}
        \end{centering}
        \caption{\label{fig:L_mpart_diff} The luminosity of the accretion disc over time, depending on the portion size of the injected matter.}
        \end{centering}
\end{figure}

In addition to the spatial resolution that depends on smoothing length, the mass resolution may also impact our accretion disc. To test this, we again performed a set of simulations with \texttt{Python} version of our code. Both the feeding radius $R_{\rm circ}=0.003$~pc and the smoothing length $h = 0.01$~pc remain constant, while the mass of a single mass portion is varied. The portion masses for these runs are chosen in proportion to the mass of the SMBH and are incremented by an order of magnitude in each run, from $M_{\rm part}/M_{\rm BH} = 10^{-7}$, up to $M_{\rm part}/M_{\rm BH} = 10^{-2}$ (the latter corresponding to the ratio in the SPH simulations). We set a target for total mass to be fed in a quater of the total run time of $500$~kyr. The target mass is the same as the mass of one portion with the ration $M_{\rm part}/M_{\rm BH} = 10^{-2}$ - $4\times10^4\msun{}$. The amount of mass contained in each component is shown in Fig. (\ref{fig:MassGrowth_mpart_diff}), where colors correspond to the component and thinner lines correspond to the simulations with less massive feeding particles. The more massive mass portions resulted in a more step-like total growth curve (black), and a spikier changes in accretion disc mass (blue), but even so, over the longer period of time the mass in each component seems to converge. In Fig. (\ref{fig:L_mpart_diff}) we show that the very sudden increases in of disc mass result in large luminosity spikes, which, while short, are also highly super-Eddington. This suggests that a limit on luminosity is necessary if very massive particles are used. At the same time, the accretion disc of the scale of used in these runs ($\sim 0.01$~pc) is probably too small for simulations that would also make use of particles as massive as used in the tests - the fact that parameters converge even in this limiting case suggests that over the longer timescales relevant to larger simulations the undesired short bursts of accretion may average out. We discuss the applicability of our model to large-scale simulations in Section \cref{sec:appl}.

In general, care should be taken when choosing the spatial resolution of the main simulation, as it could result in an artificially lower accretion disc mass and/or excessive instantaneous feeding/escape. Choosing a sink radius much larger than approximately the intended accretion disc size may also result in accretion of matter with larger than intended angular momentum and matter accumulation in the furthest parts of an unphysically large accretion disc. Our prescription does not account for the possible fragmentation and star formation at the outskirts of the disc; we leave this topic for further research.


In addition to exploring our own model, we can compare its results to the simpler two-stage accretion disc particle method (DP) using accretion data from the nFB simulation \texttt{nFBr0} as a baseline. We cannot use \texttt{FB} simulations for this because feedback has an effect on the dynamical evolution of the simulation. A convenient prescription for the rate of SMBH accretion that depends only on parameters readily available in \texttt{nFBr0} data is given by \citep{AccDiscParticle}:
\begin{equation}
\dot{M}_{\rm BH} = \rm{min}\left[ M_{\rm disc} / t_{\rm visc}, \dot{M}_{\rm Edd}\right],
\label{eq:DP_main}
\end{equation}
where $M_{\rm disc}$ is the accretion disc mass, $t_{\rm visc}$ is an arbitrarily chosen constant viscous timescale and $\dot{M}_{\rm Edd}$ is the Eddington mass accretion rate. Note that $M_{\rm disc}$ used here is not equal to the disc mass in nFB run; instead, $M_{\rm disc}$ is given by the rate of sink accretion, subtracting the mass added to the SMBH using the prescription (\ref{eq:DP_main}).

We use a set of four \texttt{Python} scripts that describe the evolution of the black hole and the gas reservoir coupled to it. In addition to the Eddington-limited model, we run scripts with no limits and a an additional limit on the maximum allowed accretion disc mass according to the gravitational stability criterion from \citep{Pringle1981}:
\begin{equation}
M_{\rm disc,max} \approx \frac{H}{R} M_{\rm BH},
\label{eq:Stable}
\end{equation}
where the disc height-radius ratio is taken to be a constant $H/R = 0.002$, typical for a thin disc. 

\begin{figure}
	\includegraphics[width=1.05\columnwidth]{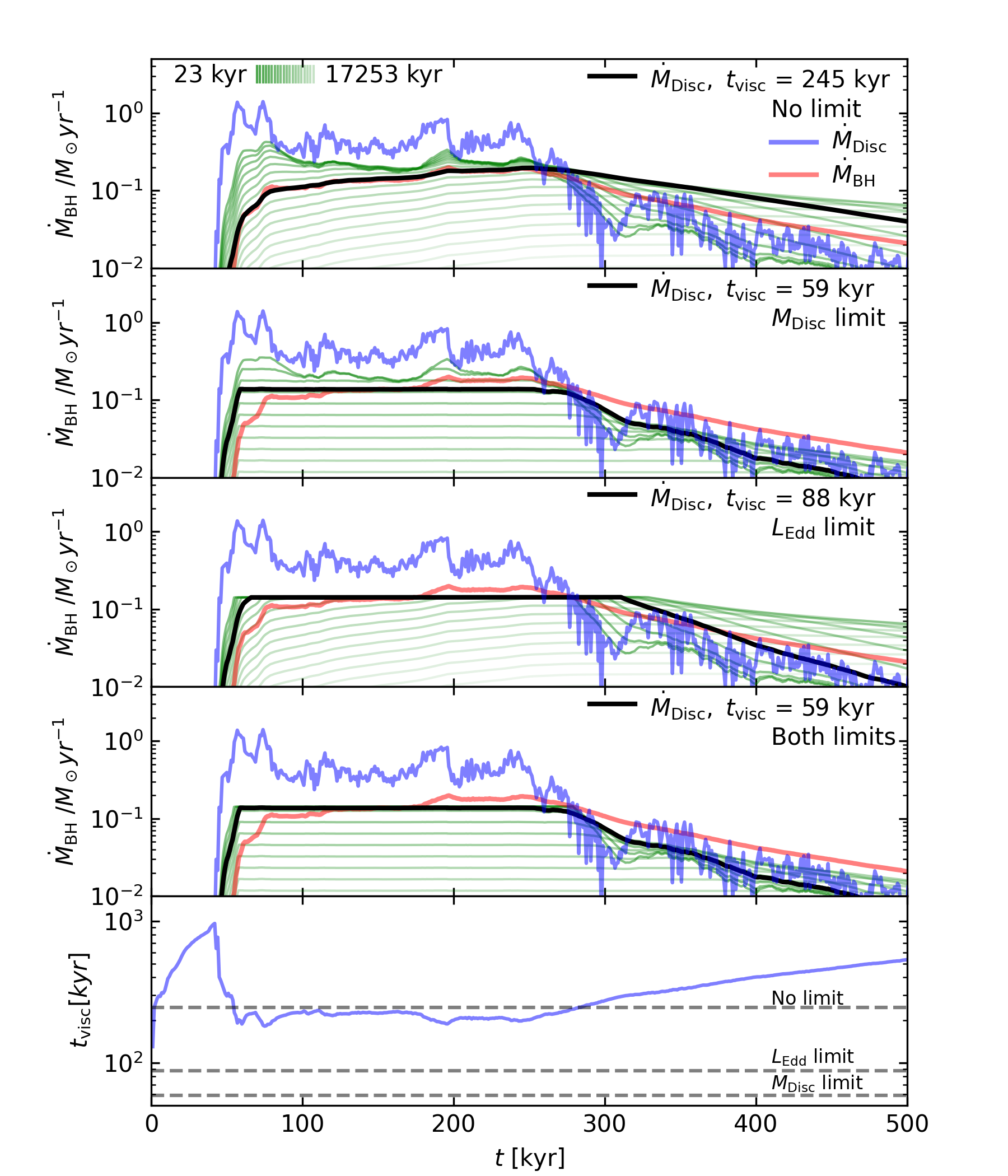}
	\vspace{-0.75cm}
    \caption{A fit of different two-stage accretion prescription to the data of the simulation \texttt{nFBr0}. Green lines represent calculations with different choices of the $t_{\rm visc}$ parameter, blue lines show the accretion on to the accretion disc particle, red - on to the SMBH using our accretion disc prescription, while the black curve is the best fitting two-stage prescription result. A plot of $t_{\rm visc} = M_{\rm disc} / \dot{M}_{\rm BH}$ is shown in the bottom panel; only the `no limit' two-stage model corresponds with the timescales found in the AD model for a period of $\sim200$~kyr.}
    \label{fig:DP_post}
\end{figure}

We show the results of this procedure in Fig. \ref{fig:DP_post}. The blue and red curves represent the rate of disc and SMBH feeding from the simulation \texttt{nFBr0}, respectively, and green ones represent the DP calculations with different choices of $t_{\rm visc}$. A disc evolution that most resembles the SMBH feeding in the \texttt{nFBr0} simulation is marked in black; when calculating it, we weigh the whole $500$~kyr duration evenly. The bottom panel shows how the closest DP calculations compare with the variable $t_{\rm visc} = M_{\rm disc} / M_{\rm \dot{M}_{\rm BH}}$ calculated from \texttt{nFBr0}. The \texttt{No Limit} approach seems to work best for the $\sim200$~kyr period where the variable $t_{\rm visc}$ remains approximately constant, while overestimating the feeding in later stages. Applying the luminosity and/or disc mass limits substantially reduces the best-fitting $t_{\rm visc}$ which also reduces the time lag between disc and SMBH feeding. We note that even if the evolution seen in the main simulation could be approximated, the `correct' $t_{\rm visc}$ is not known {\em a priori} and could be difficult to deduce as it corresponds to relatively small feeding radii $R_{\rm feed}\sim 2-5 \times10^{-4}$~pc. These values are approximately an order of magnitude lower than the $R_{\rm circ} \sim 1e-3$~pc of the majority of the accreted particles in the hydrodynamical simulations. This means that correcting only for angular momentum would not be enough to reproduce accretion disc results with AD prescription.

\subsection{Star formation on the disc outskirts}\label{sec:lost_gas} 

Over the induced activity period about half of matter added to the accretion disc particle escapes through the outer boundary (Fig. \ref{fig:MassGrowth} - green) in all of our simulations. At the moment, simulations just track the amount of matter escaping over time; it is not taken into account when calculating the evolution of the accretion disc or returned to the main hydrodynamical simulation. Neglecting this `lost mass' allows us not to complicate the calculations, but it is worthwhile to consider the possible interaction between this gas and the rest of the gas close to the SMBH as we plan to do when improving the prescription in the future. In particular, accounting for this gas in the hydrodynamical simulation may lead to either additional SMBH accretion, additional star formation in the centre, or both.

We expect that interactions between the gas escaping the accretion disc and the surrounding gas can produce additional accretion. Gas escaping via the outer boundary has $R_{\rm circ}\sim0.01$~pc, therefore needs only a small reduction in angular momentum to get accreted back, so adding this gas to the hydro simulation can result in additional accretion as it collides with infalling material.Although the majority of mass escapes the disc in the later stages of the AGN episode, the peak escape rate occurs at about the same time as central cavities appear in the FB runs, when the central disc structure is somewhat settled. Additional gas might even prevent central cavities from forming, as the total mass of gas escaping via the outer boundary is comparable to the total mass contained within the central few parsecs ($\sim10^4\,\msun{}$).

We can also envision a scenario, similar to the one described in \cite{Hobbs2011}, in which this escaping gas provides an additional barrier to accretion as the mixing of different angular momentum gas creates a peaked angular momentum distribution and a pronounced ring with $r>0.01$~pc. So the escaping matter might extend the AGN episode by providing some additional material to the accretion disc, but it might also create a more dense environment around the SMBH - maybe leading to star formation.

The main obstacle to star formation in the vicinity of the SMBH is the tidal force in the region where SMBH potential dominates ($r<0.8$~pc in our case). As a first approximation, to allow for star formation, the density in a given region has to exceed the tidal density $\rho_{\rm tidal}$:
\begin{equation}
    \rho_{\rm esc} \gtrsim \rho_{\rm tidal} \simeq  \frac{3 M_{\rm BH}}{4 \pi R^3},
    \label{eq:tidal}
\end{equation}
where $\rho_{\rm esc}=M_{\rm esc} / V$ is the density of the escaping gas, while $V$ is the volume of the region containing that gas. We can very roughly estimate where star formation might occur due to the escaped gas if we assume that all the escaped matter stays within this volume. The upper density estimate can be found by assuming that all the gas stays within a wedge-shaped disc of angle: 
\begin{equation}
    \alpha=\arctan\left( \frac{H_{\rm max}}{2R_{\rm max}}\right),
\end{equation}
here $R_{\rm max}$is the radius where the accretion disc is at its maximum height $H_{\rm max}$. We can use $\pi/2 - \alpha$ to define a cone sector of a sphere; then the angle subtended by the disc is $4\pi$ minus the angle of the conical sector. Thus the tidal density is given by:
\begin{equation}
     \rho_{\rm esc} = \frac{M_{\rm esc}}{V_{\rm sphere} - 2 V_{\rm sector}}=\frac{3M_{\rm esc}}{4\pi R^3 \cos\left(\pi/2 - \alpha\right) }.
\end{equation}

In our simulations, the value of $\alpha$ is very small, of order $10^{-3}$. In this case, $\rho_{\rm esc}$ is larger than $\rho_{\rm tidal}$ at $R<1$~pc after enough matter escapes the disc ($t=180$~kyr in FB simulations). Admittedly, this is an extreme case. Density is greatly reduced if we allow for even a small variation of the accretion disc plane. Consider a small angle $\beta$ by which the accretion disc tilts. If we assume that the disc oscillates quickly enough, we expect that this effectively results in matter distributing over a larger volume defined by a larger wedge angle $\bar{\alpha} = \alpha + \beta$. Angles as small as $\beta\gtrsim0.25\deg$ result in density low enough that star formation is no longer possible, even if all of the matter escaping the disc throughout the $500$~kyr duration of the FB simulations accumulates within the specified volume.

Since discs that allow for star formation appear to be very thin, an interesting aspect to consider is the required spatial resolution in order to be able to track star formation around the SMBH in an SPH simulation. For simplicity, consider, that the minimum resolvable height of the disc is the minimum smoothing length of an SPH particle in a given simulation: $H(R) \sim h_{\rm min}$. At radii below some threshold, where the disc is unresolved in the $z$-direction, its height remains constant and the volume can be approximated by a cylinder, giving a modified expression for $\rho_{\rm esc}$:
\begin{equation}
     \rho_{\rm esc} = \frac{M_{\rm esc}}{ \pi R^2 H}.
\end{equation}
Equating this with the tidal density we get we get the maximum height required to reach the tidal density at a given radius:
\begin{equation}
     H_{\rm max} =\frac{4}{3} \frac{M_{\rm esc}}{ M_{\rm BH}}R.
\end{equation}
In our simulations, this value is
\begin{equation}
     H_{\rm max, FB} \sim 0.013 R; \qquad H_{\rm max, nFB} \simeq 0.02 R,
\end{equation}
where the differences arises due to differences in $M_{\rm esc}$. Given that our simulations have $h_{\rm min} = 0.01$~pc, we see that they would not be able to resolve star formation closer than $R \sim 0.5-0.8$~pc from the SMBH.

\subsection{Applicability to larger scales} \label{sec:appl}

So far, we only apply our accretion disc in a system where the the spatial resolution is comparable to the size of the accretion disc, but it should be possible to apply the same approach to larger scales.  \cite{Salas2021} performed simulations of the central molecular zone (CMZ) with turbulence driving. They suggest that turbulence results in a quasi-continuous inflow of matter towards the centre; they estimate the effective viscosity $\nu$ using the $\alpha$-accretion disc theory and find it consistent with \cite{Sormani2018} models of nuclear rings present in the central $\sim1$~kpc of galaxies, where viscosity due to turbulence is included as a free parameter. This suggests that the accretion disc grid can be extended outward in a straightforward manner up to the scales of at least tens of parsecs, where circumnuclear discs and/or rings are found. But extending the disc this far may be unnecessary, since parsec-scale resolutions are common in modern galaxy simulations; therefore, extending the sub-resolution grid out to $\sim1$~pc should be sufficient. However, this would still result in a temporal and spatial resolution and would require assuming a continuous disc in the centre, which is clearly not the case in many circumstances (eg. current GC, cf. \cite{gas}, or the results of our FB simulations). In addition, an overly large sink radius would result in the smoothing out of even relatively large perturbations, leading to relatively smooth and long but weak periods of AGN activity. Some of the downsides of this extension may be circumvented by applying more stringent accretion criteria or, possibly, modifying the evolution equation itself, by accounting for disc instabilities and turbulence. Given that the current prescription requires very little computational power to process, some complication to the underlying model will not reduce its applicability from a processing power standpoint. 

Another interesting possibility is a combination our accretion disc method with a method applicable to larger scales. For example, \cite{Alcazar2016} applies and extends an analytical model of mass transport due to gravitational torques developed by \cite{Hopkins2011} to determine BH accretion in cosmological simulations:
\begin{equation}
    \dot{M}_{\rm BH} = \left(1 - \eta\right)\times\dot{M}_{\rm Torque},
\end{equation}
where $\dot{M}_{\rm Torque}$ is the mass transport due to the local properties of the unresolved central region. The model is intended to be used in cosmological and other large scale simulations; it describes the inflow of matter from scales of $0.1-1$~kpc down to the SMBH (sub-parsec) scales. It would be interesting to estimate $\dot{M}_{\rm disc}$ following the same formalism, especially in large-scale simulations with temporal resolution fine enough for the delay of feedback to have a meaningful impact.

\section{Conclusions}\label{sec:Conclusions}

We developed a simple 1D accretion disc prescription coupled to the SMBH sink particle in \texttt{Gadget-3} in order to increase the realism of black hole accretion and feedback. The prescription is based on the $\alpha$-thin accretion disc model of \cite{ShakuraSunyaev}, but uses the Paczy\'{n}ski-Wiita potential \citep{Paczy1980}. We assume that the disc is stable and quickly returns to a quasi-steady state after each mass injection. We test the prescription by simulating a retrograde collision between a torus-shaped ring surrounding a SMBH and an infalling cloud in an environment similar to the Galactic Centre. We run three sets of four simulations: with feedback from our accretion disc (FB), without feedback (nFB) and with instantaneous accretion and feedback (INST). The disruption of the initial system results in an AGN phase lasting a couple hundred kyr. Feedback reduces the total accreted mass in both sets of feedback simulations ($M_{\rm acc.tot} \sim 6\times10^4 \, \msun{}$) when compared with runs without feedback ($M_{\rm acc.tot} \sim 1.2\times10^5 \, \msun{}$) by about a half.

The major differences between simulations with instantaneous accretion and those with our accretion disc prescription are:

\begin{itemize}
\itemsep0em
    \item The growth rate of the SMBH, $\dot{M}_{\rm BH}$, is reduced and spread more evenly over time in the accretion disc prescription simulations; the change in luminosity $L_{\rm disc}$ closely follows $\dot{M}_{\rm BH}$.

    \item Radiation from the disc carries away $\eta\sim 6.25\%$ of the rest mass energy of infalling matter, which is expected in the Paczy\'{n}ski-Wiita potential and within $10\%$ of the expected value from the relativistic Schwarzschild solution.
    
    \item Feedback in the FB simulations expels gas from the central $0.1-1$~pc region, producing a central cavity. This is not reproduced in INST runs, although there the aggregate energy input into the gas is higher by a factor $\sim 1.5-2$.
    
    \item A significant amount of matter escapes via the outer boundary of the accretion disc; we neglect this in our current simulations.
    
\end{itemize}

While improvements are necessary, we show that the current implementation of the accretion disc sub-grid prescription works consistently while requiring negligible additional computational power. It provides robust results that differ significantly from instantaneous feeding prescription. Our approach is less reliant on free parameters, most importantly the viscous timescale used to artificially delay SMBH feedback. Thus, our accretion disc prescription should be especially useful in simulations of galactic nuclei on scales of tens of parsecs, where a lot of questions about the interplay between feeding and feedback and their link to star formation remain unanswered.

In future, we plan to improve the model by tracking the direction and warping of the disc plane and by consistently tracing the gas that is removed form the accretion disc via its outer boundary. Inclusion of these effects may help us to better understand how collimated feedback affects the surrounding gas.

\section*{Acknowledgements}
This research was funded by the Research Council Lithuania grant no. S-MIP-20-43. The simulations were performed on the supercomputer GALAX of the Center for Physical Sciences and Technology, Lithuania. We thank Jonas Bialopetravičius for his helpful insights concerning the \texttt{Python} implementation of the accretion disc code.

\section*{Data availability}

A \texttt{Python} implementation of the accretion disc particle is available at \href{https://github.com/Caradryan/accretiondisc}{\texttt{https://github.com/Caradryan/accretiondisc}}. Simulation results and \texttt{Gadget-3} implementation are available upon reasonable request. 




\bibliographystyle{mnras}
\bibliography{literatura} 




\appendix
\section{Thin disc evolution equations in the P-W potential} \label{App:Derivations}

For completness we provide a more detailed derivation of thin accretion disc equations, following \cite{frank_king_raine_2002} but using the \cite{Paczy1980} potential (PW). Equations (\ref{eq:mass_cons})-(\ref{eq:insert}) and (\ref{eq:mdot_par})-(\ref{eq:solve_for_c_G}) are included for posterity, as they are identical to the Keplerian case. For a more detailed description of the Keplerian case consult \cite{Pringle1981} or \cite{frank_king_raine_2002}. 

The disc is characterized by its surface density $\Sigma(R, t)$, which is given by integrating the gas density $\rho$ in the $z$~direction. the amount of matter contained in a single annulus between $R$ and  $R + \Delta R$ is $2\pi R\Delta R\Sigma$; similarly, the total angular momentum is $2\pi R\Delta R\Sigma R^2 \Omega$. The rate of change of these quantities is determined by the net flow from neighbouring annuli: 
\begin{equation}
\begin{split}
\frac{\partial}{\partial t}(2 \pi R \Delta R \Sigma) &= v_R(R, t) 2 \pi R\Sigma (R, t) \\ &- v_R(R+\Delta R, t) 2 \pi (R + \Delta R) \Sigma(R+\Delta R, t) \\ 
&\approx -2 \pi \Delta R \frac{\partial}{\partial R} (R \Sigma v_{R}).
\label{eq:mass_cons}
\end{split}
\end{equation}
As $\Delta R \rightarrow{} 0$, we get the mass conservation equation:
\begin{equation}
    R \frac{\partial \Sigma}{\partial t} + \frac{\partial}{\partial R} \left( R \Sigma v_{R} \right) = 0.
\end{equation}

The conservation of angular momentum is constructed in the same way from the rate of change of angular momentum, but an additional transport term due to the viscous torques $G(R, t)$ is included:
\begin{equation}
\begin{split}
\frac{\partial}{\partial t}(2 \pi R \Delta R \Sigma R^2 \Omega) &= v_R(R, t) 2 \pi R\Sigma (R, t) R^2 \Omega(R) \\ &- v_R(R+\Delta R, t) 2 \pi (R + \Delta R) \Sigma(R+\Delta R, t) \\ 
&\times(R+\Delta)^2\Omega(R+\Delta R) +\frac{\partial G}{\partial R}\Delta R \\
&\approx -2 \pi \Delta R \frac{\partial}{\partial R} (R \Sigma v_{R} R^2 \Omega)+\frac{\partial G}{\partial R}\Delta R.
\label{eq:mom_cons}
\end{split}
\end{equation}
again taking $\Delta R \rightarrow{} 0$ we arrive at the angular momentum conservation equation:
\begin{equation}
R \frac{\partial}{\partial t}(\Sigma R^2 \Omega) + \frac{\partial}{\partial R}(R \Sigma v_R R^2\Omega) = \frac{1}{2\pi}\frac{\partial G}{\partial R}.
\end{equation}

The expression for the torque includes the viscosity term $\nu=\alpha c_{\rm{s}} H$:
\begin{equation}
G(R) = 2\pi R \nu \Sigma R^2 \frac{\partial \Omega}{\partial R}.
\label{eq:torq}
\end{equation}
Using (\ref{eq:mass_cons}) we can simplify (\ref{eq:mom_cons}):
\begin{equation}
    R \Sigma v_R \frac{\partial}{\partial R}\left(R^2 \Omega\right) = \frac{1}{2\pi}\frac{\partial G}{\partial R},
    \label{eq:simpl}
\end{equation}
note that we assume that the $\partial \Omega / \partial t$ term can be safely neglected as the change in potential due to the increase in SMBH mass is negligible.

Combining (\ref{eq:mass_cons}) and (\ref{eq:simpl}) allows us to eliminate $v_R$.
\begin{equation}
R \frac{\partial \Sigma}{\partial t} = -\frac{\partial}{\partial R}\left(R \Sigma v_R\right) = -\frac{\partial}{\partial R} \left[\frac{1}{2\pi \frac{\partial}{\partial R}\left(R^2 \Omega\right)} \frac{\partial G}{\partial R} \right].
 \label{eq:insert}
\end{equation}

We now introduce the PW potential
\begin{equation}
    \phi = -\frac{\rm{G}M_{\rm BH}}{R-R_g}, 
\end{equation}
from which a modified expression of angular velocity $\Omega$ follows:
\begin{equation}
    \Omega = \left(\frac{1}{R}\frac{{\rm d}\phi}{{\rm d}R}\right)^{1/2} = \left( \frac{\rm{G}M_{\rm BH}}{R^3} \right)^{1/2} \left( \frac{R}{R-R_g}  \right).
\end{equation}
Here $R_g$ is the Schwarzschild radius. 

\begin{figure}
	\includegraphics[width=\columnwidth]{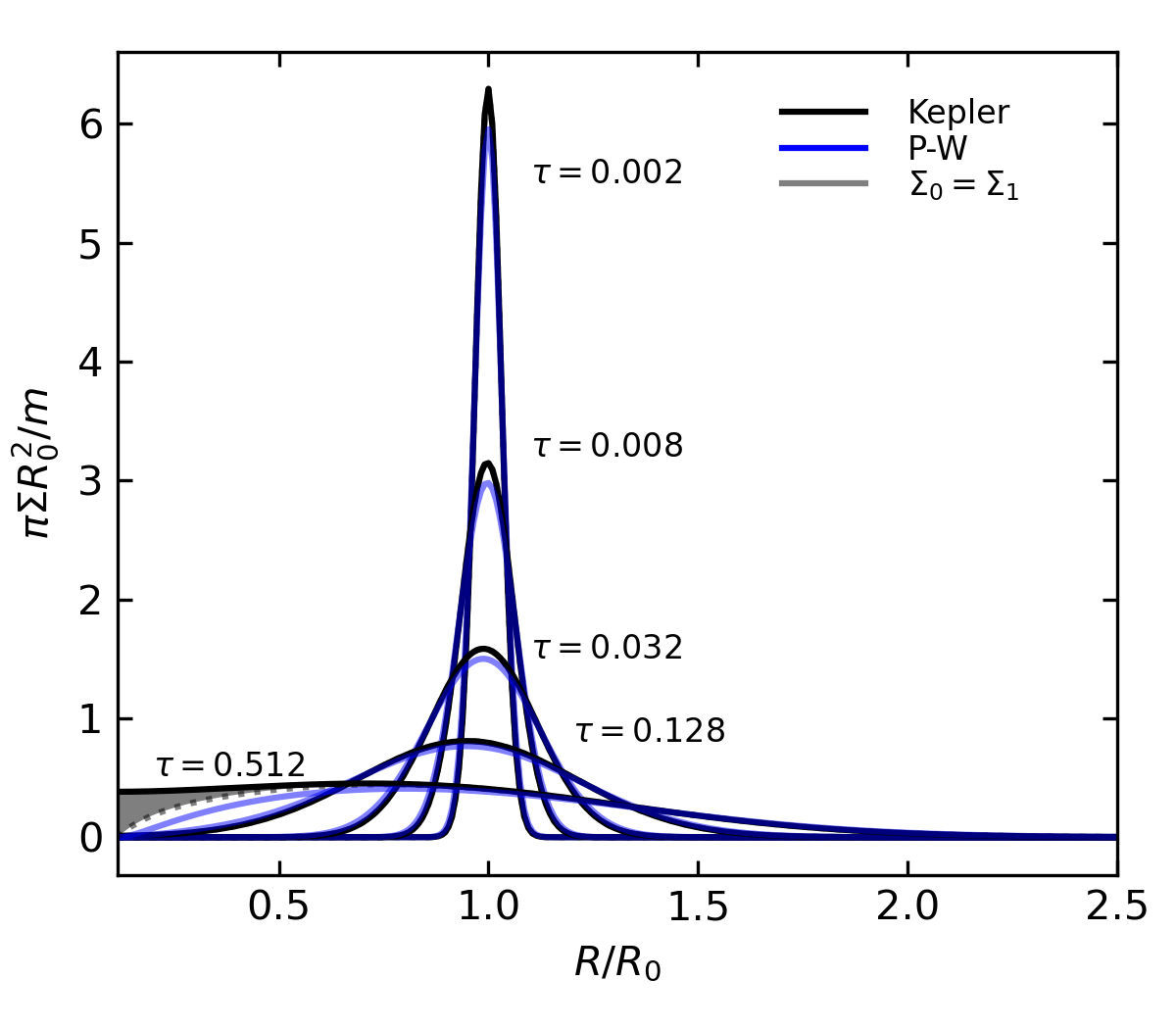}
	\vspace{-0.75cm}
    \caption{Diffusion evolution test. The blue curves correspond to evolution with eq. (\ref{eq:Diff_again}), where diffusion appears to be slightly more rapid. Black lines correspond to the standard evolution equation (to which (\ref{eq:Diff_again}) reduces with $R_{g}=0$).  $R_0 = 3\times10^{-3}$~pc and $\tau = 12 \nu t / R_0^2$ is the dimensionless time, where $\nu$ a constant \citep{Pringle1981}. }
    \label{fig:DiffTest}
\end{figure}

Inserting expressions for 
\begin{equation}
    \frac{\partial}{\partial R}\left(R^2 \Omega\right) = \frac{R}{2}\frac{R - 3R_g}{ \left(R-R_{\rm g}\right)^2}\left( \frac{G M}{R^3} \right)^{1/2}
\end{equation}
and 
\begin{equation}
\begin{split}
    \frac{\partial G}{\partial R} &=\frac{\partial }{\partial R}2\pi R \nu \Sigma R^2 \Omega' \\
    &= 2 \frac{3}{2} \pi R \nu \Sigma R^2 \left( \frac{G M}{R^3} \right)^{2} \left( \frac{R - R_{\rm g}/3}{\left(R-R_{\rm g}\right)^{1/2}}  \right)
\end{split}
\end{equation}
to (\ref{eq:insert}) and simpolifying we arrive at the main diffusion equation: 
\begin{equation}
\frac{\partial \Sigma}{\partial t} = \frac{3}{R}\frac{\partial }{\partial R} \left[ \frac{\left(R-R_{\rm g}\right)^2}{R^{1/2}\left(R-3R_{\rm g}\right)}\
 \frac{\partial}{\partial R} \left ( \nu \Sigma R^{3/2} \frac{R-R_{\rm g}/3}{\left(R-R_{\rm g}\right)^2} \right )  \right];
 \label{eq:Diff_again}
\end{equation}
This is the main viscous evolution equation with the PW potential that our prescription solves by finite differences. Equation (\ref{eq:Diff_again}) reduces to the standard Keplerian form if we set $R_{\rm g}=0$. So see whether it behaves as expected we perform a diffusion evolution test with constant viscosity. Results for both PW (blue) and standard Keplerian (black) potentials are shown in Fig. (\ref{fig:DiffTest}). We see a noticeable effect of the boundary condition $\Sigma_[0] = 0$ and a slight difference between the PW and Keplerian distributions, in that the PW potential results in somewhat faster diffusion. This is expected, because the PW potential effectively brings the material `closer' to the origin of the potential, so its evolution is faster.

\begin{figure}
	\includegraphics[width=\columnwidth]{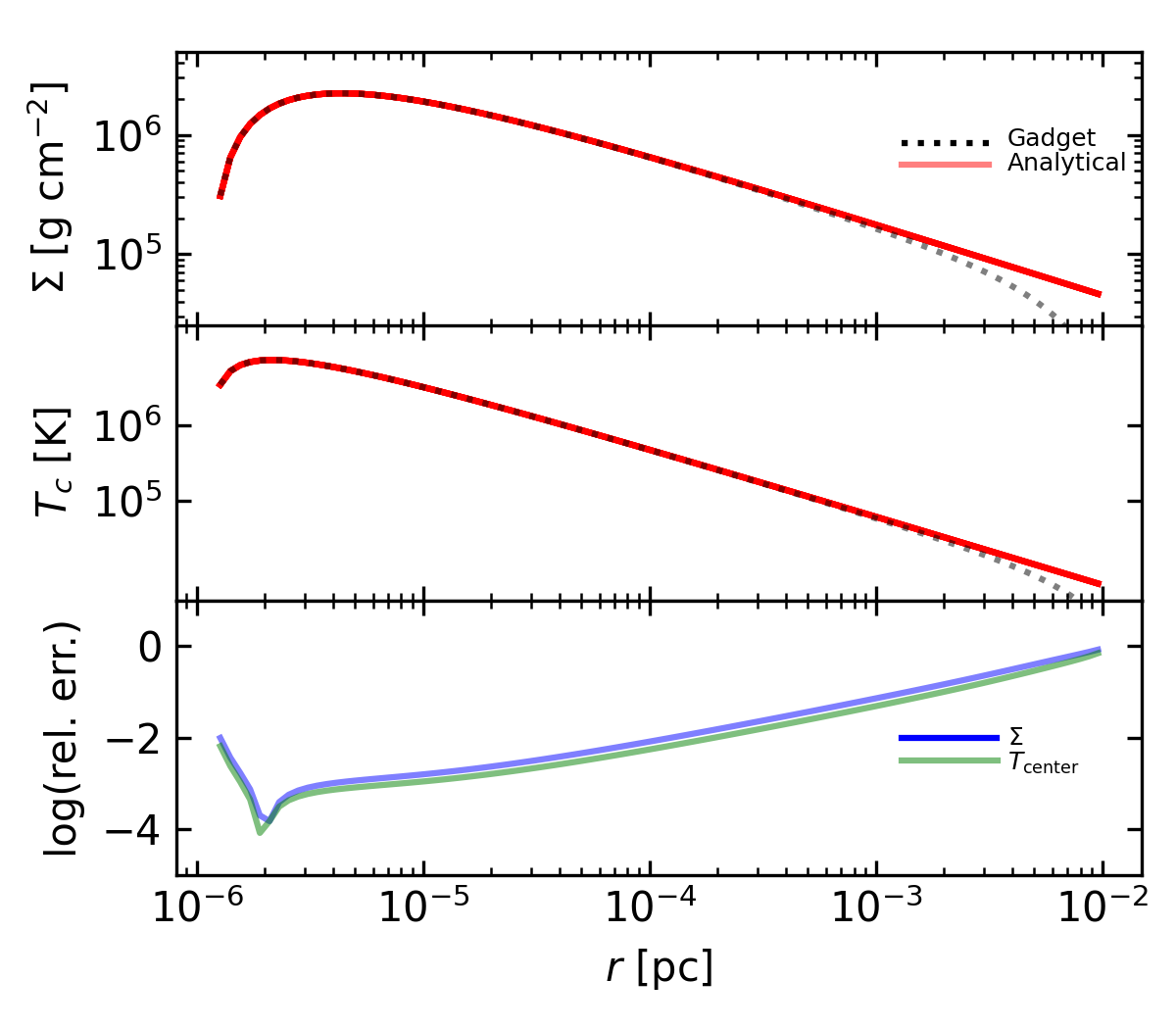}
	\vspace{-0.75cm}
    \caption{ A comparison between the expected analytical (red) and model (dotted black) radial structure of the accretion disc. The bottom panel shows the relative deviation from the analytical solution ($\log| \Sigma_{\rm model} / \Sigma_{\rm expected} -1|$ in blue and same for $T_{\rm center}$ in green). Deviations are larger closer to the disc boundaries.}
    \label{fig:CompareAnal}
\end{figure}

We assume that the disc is stable, thin and its parameters should tend to solutions for the steadily accreting disc. In a sense, each feeding cycle is a perturbation, while viscous diffusion distributes the matter to ever more closely resemble the ideal quasi-steady thin accretion disc. This assumption greatly simplifies the calculation of the disc parameters, allowing us to disregard the time derivative in the conservation equations (\ref{eq:mass_cons}) and (\ref{eq:mom_cons}). From these we get:
\begin{equation}
\begin{split}
    \dot{M} &= 2\pi R \Sigma (-v_R),
    \label{eq:mdot_par}
\end{split}
\end{equation}
where $\dot{M}$ is the accretion rate, and
\begin{equation}
\begin{split}
   R \Sigma v_R R^2 \Omega &= \frac{G}{2\pi} + \frac{C}{2\pi}. 
   \label{eq:solve_for_c}
\end{split}
\end{equation}
Here C is some constant. Inserting the expression (\ref{eq:torq}) for $G$ we get
\begin{equation}
\begin{split}
   -\nu \Sigma \frac{\partial \Omega}{\partial R} &= \Sigma (-v_R)\Omega + \frac{C}{2 \pi R^3}.
   \label{eq:solve_for_c_G}
\end{split}
\end{equation}
Equation (\ref{eq:solve_for_c_G}) can be solved for $C$; applying $\partial \Omega/\partial R=0$ at the innermost stable orbit ($R=3R_{\rm g}$) we get:
\begin{equation}
\begin{split}
   C &= -\frac{3}{2}\dot{M}(GM\cdot3 R_g)^{1/2}.
\end{split}
\end{equation}
This allows us to get a useful expression:
\begin{equation}
\begin{split}
   \nu \Sigma &= \frac{\dot{M}}{3\pi} \left( \frac{1}{R-R_{\rm g}} - \frac{3^{3/2}R_g^{1/2}}{2 R^{3/2}}  \right) \left( \frac{ \left(R - R_{\rm g}\right)^2 }{R - R_{\rm g}/3} \right)
\end{split}
\end{equation}
Using this we can get the expression for energy dissipation $D(R)$
\begin{equation}
\begin{split}
   D\left(R\right) &= \frac{R^2}{2} \nu \Sigma \left(\frac{\partial \Omega}{\partial R}\right)^2 \\ &= \frac{3}{8}\frac{\dot{M}}{\pi}\frac{GM}{R}\left( \frac{1}{R-R_{\rm g}} - \frac{3^{3/2} R_{\rm g}^{1/2} }{2 R^{3/2} }  \right) \left( \frac{ (R - R_{\rm g})^2 }{R - R_{\rm g}/3} \right)
\end{split}
\end{equation}
Using this we can get then expression for central temperature
\begin{equation}
\begin{split}
   T_c^4 &= \frac{3\tau}{4\sigma}D(R)\\
   &=\frac{27}{32}\frac{\tau}{\sigma}\nu \Sigma \frac{GM}{R} \frac{(R-R_{\rm g}/3)^2}{(R-R_{\rm g})^4}, 
\end{split}
\end{equation}
where $\tau = \kappa\Sigma/2$ and $\kappa = 0.348$~cm$^2\,$~g$^{-1}$.

To get the height of the accretion disc, we again repeat the considerations outlined in \citep{frank_king_raine_2002}. Taking the Euler equation:
\begin{equation}
 \rho \frac{\partial \textbf{v}}{\partial t} + \rho \textbf{v}\cdot \Delta \textbf{v} = -\Delta P + \textbf{f}_g, 
\end{equation}
where $\textbf{f}_g,$ is the force density due to gravitation, we assume hydrostatic equilibrium in the $z$ direction and neglect the velocity terms arriving at:
\begin{equation}
\begin{split}
 \frac{\partial}{\partial z} P = \rho \frac{\partial}{\partial z} (\phi) = \rho \frac{\partial}{\partial z} \left (- \frac{G M}{(R^2 + z^2)^{1/2} - R_{\rm g}} \right) 
\end{split}
\end{equation}
completing the partial derivative on the right hand side and moving $\rho$ to the left we get
\begin{equation}
\begin{split}
   \frac{1}{\rho} \frac{\partial}{\partial z} P = \frac{G M z}{(R^2 + z^2)^{1/2}\left[  (R^2 + z^2)^{1/2} - R_{\rm g} \right]^2}.
\end{split}
\end{equation}
Following the argumentation from \cite{frank_king_raine_2002}: if the scaleheight in $z$ direction is $H$, then $\frac{\partial P}{\partial z}\sim \frac{P}{H}$ and $z\sim H$. The thin disc assumption gives $H \ll R$; using $P \sim \rho c_{s}^2$ we get our final expression for $H$:
\begin{equation}
\begin{split}
   H &= c_s \left(R-R_{\rm g}\right)\left( \frac{R}{GM} \right)^{1/2},
   \label{eq:H}
\end{split}
\end{equation} 
where the speed of sound $c_{s}$ is given by:
\begin{equation}
\begin{split}
   c_s = T_{\rm c}^4 \left( \frac{\rm{k}\Gamma}{m_{\rm p}  \mu} \right)^{1/2},
\end{split}
\end{equation}
with the mean molecular weight $\mu = 0.63$ and $\Gamma = 5/3$.

To check if our equations give us the expected results we perform a test, where an accretion disc is fed at a constant rate. After some time, the disc approaches a steady state, that is, the disc's mass remains constant as does the rate of SMBH feeding and matter escaping via outer boundary. To find the analytically expected result for comparison, we solve equations for the surface density $\Sigma$:
\begin{multline}
    \Sigma^5  =  \frac{64}{729}\frac{GM\sigma_{SB}}{R\kappa} \left(\frac{\dot{M}}{\pi} \right)^3 \left( \frac{1}{R-R_{\rm g}} - \frac{3^{3/2} R_{\rm g}^{1/2}}{2 R^{3/2}}\right)^3 \cdot \\ 
     \cdot \left( \frac{3  m_{\rm p} \mu }{5 \alpha k (R-R_{\rm g})} \right)^4 \left( \frac{(R-R_{\rm g})^2}{R - R_{\rm g}/3} \right)^5,
     \label{eq:SA}
\end{multline}
where $m_{\rm p}$ is the proton mass and $k$ is the Boltzmann constant, and the central temperature $T_{\rm c}$:
\begin{equation}
    T_{\rm c}^4 = \frac{9}{64} \frac{\kappa \Sigma}{\sigma_{\rm SB}} \frac{\dot{M}}{\pi} \frac{G M_{\rm BH}}{R} 
    \left( \frac{1}{R- R_{\rm g}} - \frac{3^{3/2} R_{\rm g}^{1/2} }{2 R^{3/2}} \right)
    \left( \frac{R - R_{\rm g}/3}{ (R - R_{\rm g})^2}  \right).
    \label{eq:TcA}
\end{equation}
We input the SMBH mass $M_{\rm BH}$ and SMBH accretion rate $\dot{M}$ from the model into equations (\ref{eq:SA}) and (\ref{eq:TcA}) to get the radial structure of a steadily accreting disc as expected for the given parameters and plot it together with the radial structure obtained via model calculations in Fig. \ref{fig:CompareAnal}. The upper two panels show surface density $\Sigma$ and central temperature $T_{\rm c}$, respectively. The red curves represent the analytical results for the given disc parameters, while the dotted line shows results taken from a \texttt{Gadget} simulation. The bottom panel shows the relative deviation from the analytical solution for the both surface density (blue) and the central temperature (green). We can see, that both the central temperature and the surface density deviate more from the analytical solution the closer they get to the disc boundaries, but the agreement is generally very good; even the relatively large deviation at the outer boundary is relatively unimportant as it generates a negligible portion of the whole luminosity.

\section{Representative surface density maps} \label{app:grids}

\begin{figure*}
    \centering
    \includegraphics[width=\textwidth]{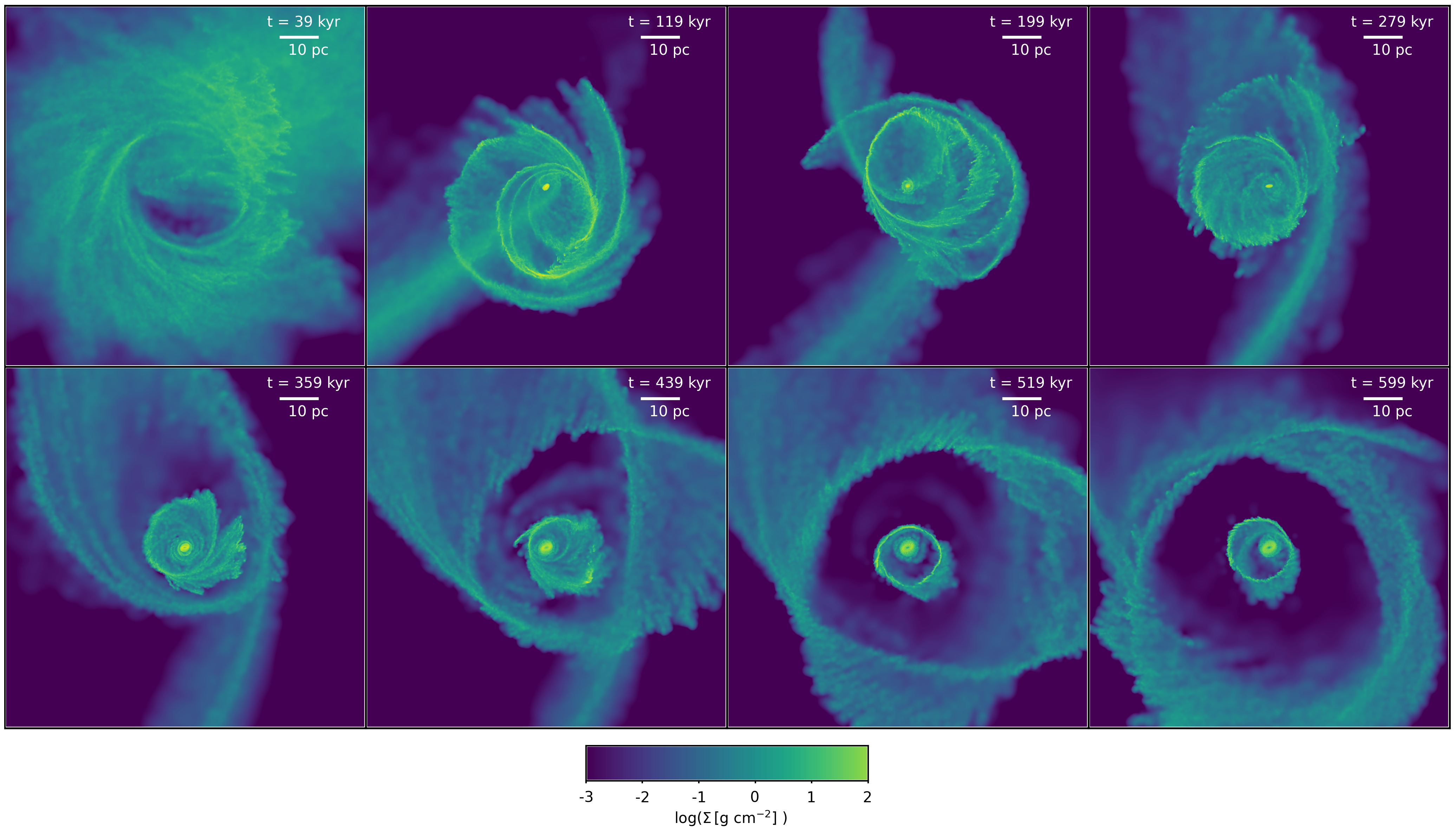}
    \caption{\label{fig:grid_00} Surface density maps of the nFB run  \texttt{nFBr0}. The change in morphology resulting from the collision as shown here is representative of all nFB runs.}
\end{figure*}

\begin{figure*}
    \centering
    \includegraphics[width=\textwidth]{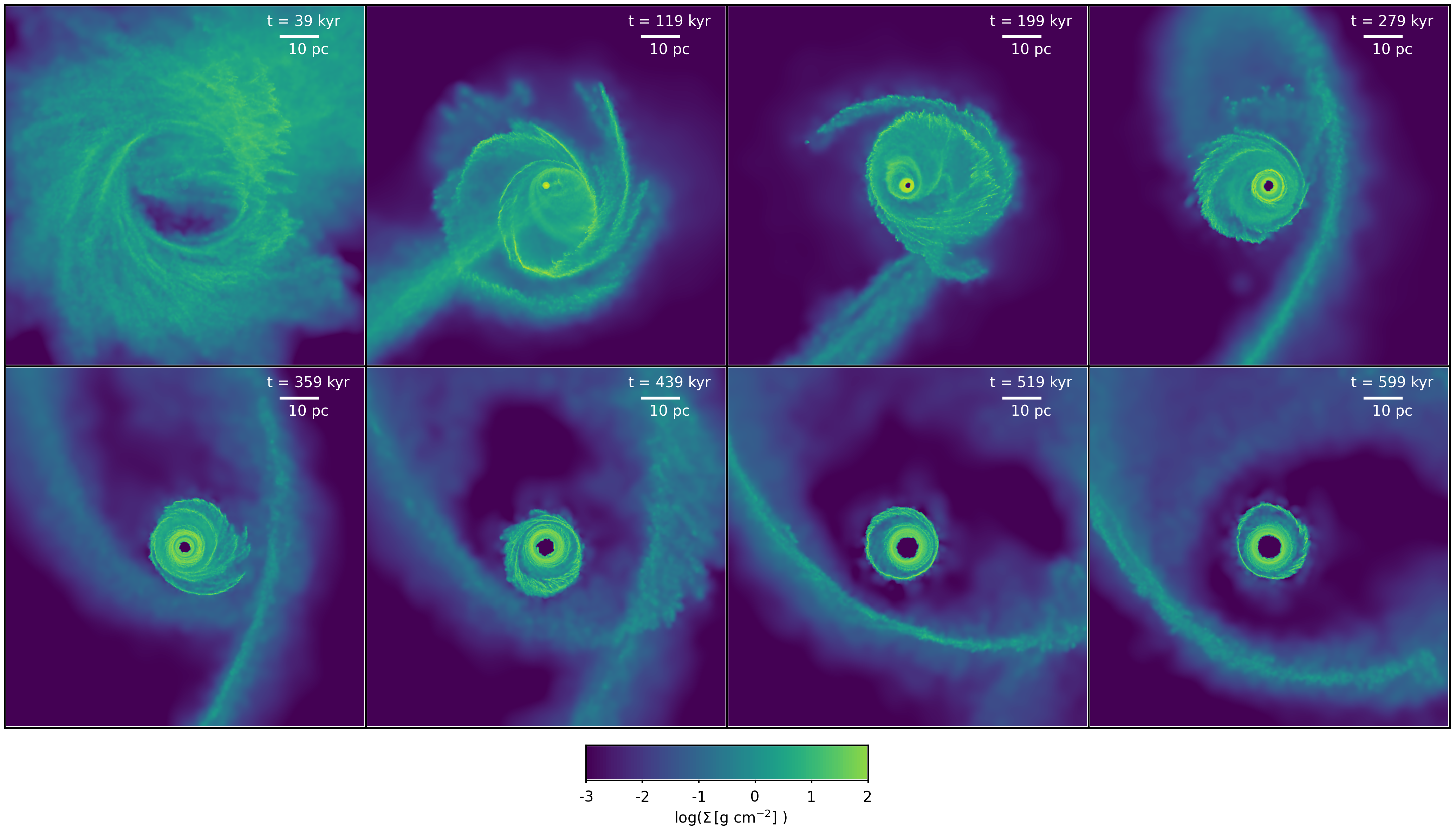}
    \caption{\label{fig:grid_10}  Surface density maps of the FB run  \texttt{FBr0}. The change in morphology resulting from the collision as shown here is representative of all FB runs except \texttt{FBr2}.}
\end{figure*}

\begin{figure*}
    \centering
    \includegraphics[width=\textwidth]{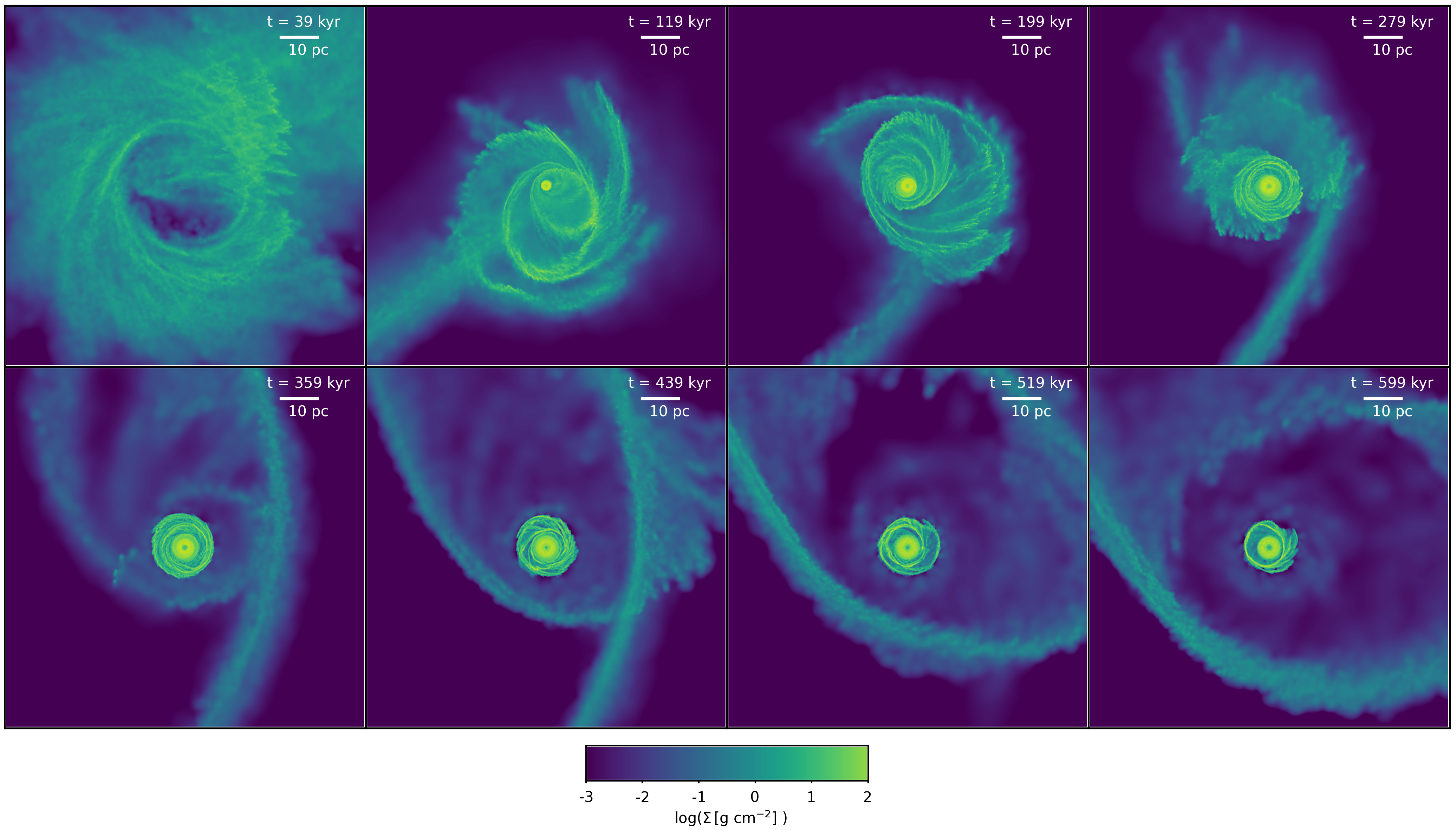}
    \caption{\label{fig:grid_00i} Surface density maps of the INST run  \texttt{INSTr0}. }
\end{figure*}

\begin{figure*}
    \centering
    \includegraphics[width=\textwidth]{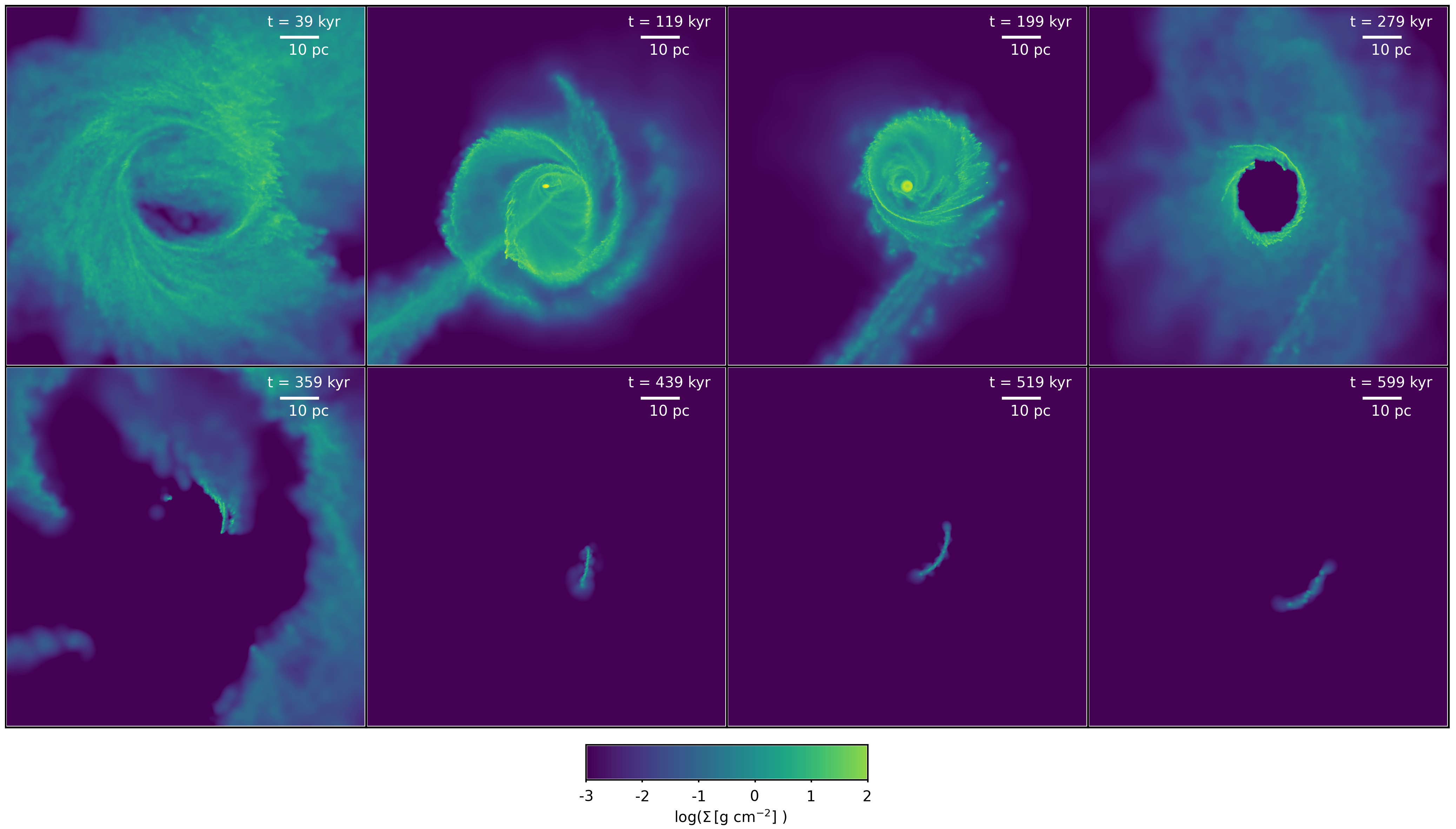}
    \caption{\label{fig:grid_12} Surface density maps of the FB run  \texttt{FBr2}. }
\end{figure*}

The following figures show density maps of four simulations at a few different stages in their evolution. In Fig. \ref{fig:grid_00} and Fig. \ref{fig:grid_10} density maps of runs nFBr0 and FBr0 are shown. We can see a very similar initial evolution. From the third panel on, we can see that the FBr0 has a distinct central cavity that increases in size as the simulation progresses. In Fig. \ref{fig:grid_00i} we see that in a run with an instantaneous feedback prescription, INSTr0, similar initial evolution occurs, but despite generally stronger feedback the run fails to produce a central cavity. Three of the runs shown, nFBr0, FBr0 and INSTr0 have the same initial particle distribution. Fig. \ref{fig:grid_12} shows the evolution of the outlier run FBr2 in which almost all of the initial gas is pushed out during the AGN phase.


\bsp	
\label{lastpage}
\end{document}